\documentclass[lettersize,journal]{IEEEtran}

\usepackage{cite}
\usepackage{amsmath,amssymb,amsfonts}
\usepackage{mathtools}
\usepackage{graphicx}
\usepackage{textcomp}
\usepackage[table]{xcolor}
\usepackage{enumitem}
\usepackage[justification=centering]{caption}
\graphicspath{{Figures/}}
\usepackage{setspace}
\usepackage{bbm}
\usepackage{booktabs}
\usepackage{multirow}
\usepackage{tabularx}
\usepackage{placeins}
\usepackage{pgfplots}
\usepackage{pgfplotstable}
\usepgfplotslibrary{groupplots}
\pgfplotsset{compat=1.15}
\pgfplotsset{every tick label/.append style={font=\tiny}}

\usepackage{hyperref}
\hypersetup{
	colorlinks=true,
	linkcolor=blue,
	filecolor=blue,
	citecolor=blue,
	urlcolor=blue,
	pdftitle={Multi-Tag Collision Recovery in UHF-RFID Using Self-Attention Decoding},
	pdfpagemode=FullScreen,
}
\usepackage{tikz}
\usepackage{dblfloatfix} 
\usetikzlibrary{positioning}
\usetikzlibrary{positioning, shapes.geometric}
\usetikzlibrary{positioning, arrows.meta}
\usetikzlibrary{decorations.pathreplacing}
\usetikzlibrary{calc, decorations.pathmorphing}
\usetikzlibrary{backgrounds}

\newcommand{\E}{\mathbb{E}}
\newcommand{\CN}{\mathcal{CN}}
\newcommand{\R}{\mathbb{R}}
\newcommand{\C}{\mathbb{C}}
\DeclareMathOperator*{\argmin}{arg\,min}

\definecolor{figcolor1}{HTML}{FF1F5B}
\definecolor{figcolor2}{HTML}{009ADE}
\definecolor{figcolor3}{HTML}{00CD6C}
\definecolor{figcolor4}{HTML}{AF58BA}
\definecolor{figcolor5}{HTML}{F28522}
\definecolor{figcolor6}{HTML}{089099}
\definecolor{convcol}{HTML}{A6761D}   

\definecolor{tblhead}{RGB}{31,59,102}   
\definecolor{tblzebra}{RGB}{244,246,249} 
\newcommand{\hd}[1]{\textcolor{white}{\textbf{#1}}} 

\tikzset{optarrow/.style={->,>={Stealth[length=4pt]},gray!65,line width=0.5pt}}

\definecolor{readercol}{RGB}{231,98,95}   
\definecolor{tagcol}{RGB}{78,135,190}     
\definecolor{fwdcol}{RGB}{45,110,185}     
\definecolor{bwdcol}{RGB}{222,52,48}      
\definecolor{leakcol}{RGB}{72,168,82}     

\begin{document}

\title{Multi-Tag Collision Recovery in UHF-RFID Using Self-Attention Decoding}

\author{\IEEEauthorblockN{Talha Aky{\i}ld{\i}z, Siva Aditya Gooty, Hessam Mahdavifar, and Najme Ebrahimi\\}
	\thanks{T. Aky{\i}ld{\i}z is with the EECS Dept., University of Michigan, Ann Arbor, MI, 48104, USA (email: akyildiz@umich.edu).}
	\thanks{S. A. Gooty is with the ECE Dept., Northeastern University, Boston, MA, 02115, USA (email: gooty.s@northeastern.edu).}
	\thanks{H. Mahdavifar is with the EECS Dept., University of Michigan, Ann Arbor, MI, 48104, USA and ECE Dept., Northeastern University, Boston, MA, 02115, USA (email: h.mahdavifar@northeastern.edu).}
	\thanks{N. Ebrahimi is with the ECE Dept., Northeastern University, Boston, MA, 02115, USA (email: n.ebrahimi@northeastern.edu).}
	\thanks{Part of this work was presented at the 2022 IEEE International Conference on RFID \cite{akyildiz2022mlaided}.}
}

\maketitle

\begin{abstract}
Passive ultra high frequency (UHF) radio frequency identification (RFID) enables battery-free tags to communicate with a reader through backscatter. When multiple tags respond in the same time slot, their waveforms overlap at the reader, and a conventional reader that follows framed slotted ALOHA (FSA) discards the resulting collided slot. This limits the throughput of the overall protocol even though the received signal still contains recoverable information about the responding tags. To address this limitation, we propose Self-Attention Tag Recovery (SATR), a transformer-based decoding algorithm that operates directly on the baseband in-phase and quadrature (I/Q) samples received during a standard tag response. SATR uses self-attention to model the temporal structure of the modulated waveform and learns candidate tag representations. It jointly estimates the number of responding tags and, more importantly, decodes the bit sequence of each detected tag. We numerically evaluate the decoding and throughput performance of SATR over a range of collision sizes and recovery configurations, and validate it with measurements of commercial UHF-RFID tags. The results show that, with proper design and training, SATR can reliably decode collisions of up to four tags. It achieves a throughput of approximately $0.815$ tags per slot under single acknowledgment and $1.87$ tags per slot under full recovery, corresponding to $2.2$ and $5.1$ times the conventional FSA limit of $1/e \approx 0.368$ tags per slot, while approaching optimal decoding performance and outperforming existing collision recovery methods.
\end{abstract}


\section{Introduction}
\label{sec:introduction}

Radio frequency identification (RFID) is a widely used automatic identification technology that supports the tracking and management of physical objects in modern Internet of Things (IoT) deployments \cite{want2006introduction,finkenzeller2010rfid,miorandi2012internet}. Typical applications include logistics, retail inventory, healthcare asset management, manufacturing, and access control \cite{sarac2010literature}. For item-level tagging, passive ultra high frequency (UHF) RFID operating in the 860--960~MHz band is commonly used due to its practical balance of read range, cost, and data rate. Passive UHF tags carry no onboard power source and instead harvest energy from the continuous wave (CW) signal radiated by the reader, responding through backscatter modulation \cite{nikitin2007differential,kim2003measurements}. This battery-free operation supports low-cost deployment at large scale, but reliably decoding tag responses becomes more difficult as the number of tags grows.

\nocite{global2008epc,schoute1983dynamic,vogt2002efficient,floerkemeier2006bayesian,knerr2010slot,vales2011multiframe,chen2014unified,shen2009separation,khasgiwale2009extracting,yu2008anti,mindikoglu2008separation,angerer2010single,angerer2010rfid,kaitovic2011rfid,kaitovic2012channel,kaitovic2013smart,fyhn2011multipacket,kang2011decoding,bletsas2012single,kargas2015fullycoherent,tan2016collision,mahdavifar2015coding,ou2015come,jin2019fliptracer,jin2018parallelbackscatter,zhang2020semi,wu2021fsvd,salah2024matrixpencil,pirayesh2023mreader,zeng2024fastmc,siala2025antennaarray,alfayoumi2025nextgen,oshea2017introduction,ye2018power,samuel2019learning,akyildiz2022mlaided}

\begin{table*}[t]
\centering
\caption{Comparison of representative UHF-RFID collision recovery methods and operating assumptions}
\label{tab:comparison}
\footnotesize
\renewcommand{\arraystretch}{1.25}
\rowcolors{2}{white}{tblzebra}
\begin{tabular}{>{\raggedright\arraybackslash}p{2.3cm} >{\raggedright\arraybackslash}p{3.3cm} >{\raggedright\arraybackslash}p{1.8cm} >{\raggedright\arraybackslash}p{2.0cm} >{\raggedright\arraybackslash}p{5cm} >{\centering\arraybackslash}p{1.1cm}}
\rowcolor{tblhead}
\hd{Method} & \hd{Main mechanism} & \hd{Extra requirement} & \hd{Processing} & \hd{Assumption or scope} & \hd{Reported tags} \\
ZF/SIC \cite{angerer2010single,angerer2010rfid} & Channel estimation and interference cancellation & Preamble & Per slot estimation & Single antenna recovery targets two tags, while arrays are used for higher orders & 2 \\
Coherent FM0 \cite{bletsas2012single} & Coherent FM0 sequence detection & Signal levels & Per slot detection & Detector is defined for two tags & 2 \\
Matrix Pencil \cite{salah2024matrixpencil} & Matrix Pencil parameter estimation & Preamble and antenna array & Per slot parameter estimation & Short FM0 preamble constrains reliable estimation beyond three tags & 2--3 \\
mReader \cite{pirayesh2023mreader} & MU-MIMO beamforming & Antenna array & Beamforming and channel calibration & Requires multiple reader antennas and was demonstrated for two tags & 2 \\
BiGroup \cite{ou2015come} & I/Q clustering with time-domain bipartite grouping & None at tag & Clustering and bit boundary extraction & Uses tag asynchronism and separable I/Q clusters. Reported recovery of all tags decreases for larger collision sizes. & 5 \\
I/Q geometry \cite{alfayoumi2025nextgen} & Clustering and Procrustes constellation alignment & Preamble anchor & Per slot optimization & Requires observable constellation clusters. FM0 decoding is validated for three tags, while support for four tags is reported. & 4 \\
\rowcolor{figcolor3!15}
\textbf{SATR (this work)} & \textbf{Self-attention with learned tag representations and FM0-aware bit decoding} & \textbf{None at tag} & \textbf{Joint learned decoding} & \textbf{Requires offline training and coverage of expected channel and collision conditions} & \textbf{4} \\
\end{tabular}
\vspace{-0.4cm}
\end{table*}

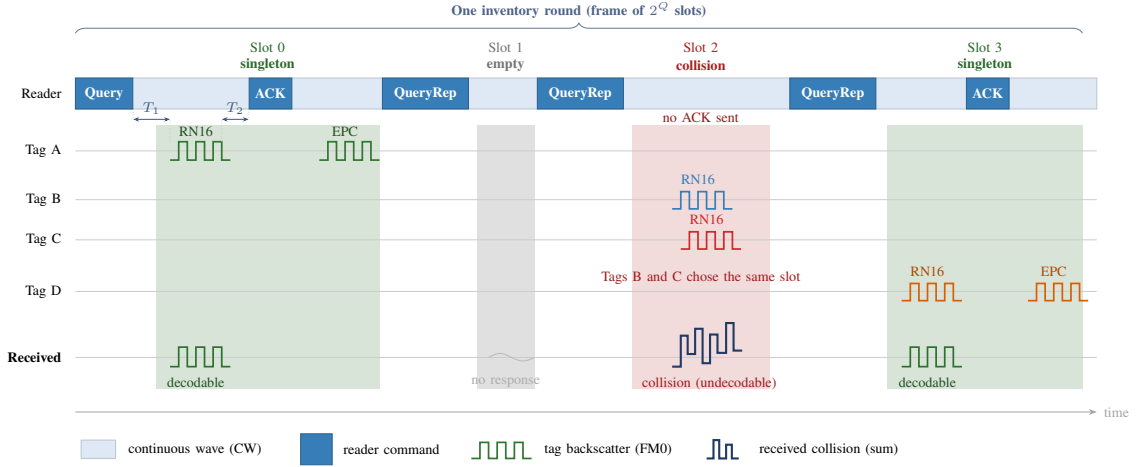
\begin{figure*}[t]
\centering
\resizebox{.9\textwidth}{!}{\definecolor{figblue}{RGB}{55,126,184}
\definecolor{navy}{RGB}{31,59,102}
\definecolor{figgreen}{RGB}{60,150,60}
\definecolor{figred}{RGB}{210,40,40}
\definecolor{figorange}{RGB}{217,95,2}
\definecolor{cwc}{RGB}{223,233,246}
\colorlet{greenlab}{figgreen!70!black}
\colorlet{redlab}{figred!80!black}
\colorlet{graylab}{black!55}
\colorlet{okband}{greenlab!18}
\colorlet{collband}{redlab!16}
\colorlet{emptyband}{graylab!22}

\newcommand{\fm}[4]{\draw[#3,line width=0.85pt]
 (#1,#2)-- ++(0.15,0)-- ++(0,#4)-- ++(0.15,0)-- ++(0,-#4)-- ++(0.15,0)-- ++(0,#4)
 -- ++(0.15,0)-- ++(0,-#4)-- ++(0.15,0)-- ++(0,#4)-- ++(0.15,0)-- ++(0,-#4)-- ++(0.15,0);}

\begin{tikzpicture}[font=\small,
  cmd/.style={fill=figblue, draw=figblue!75!black, text=white, minimum height=0.54cm, inner sep=3pt, font=\scriptsize\bfseries},
  row/.style={anchor=east, font=\scriptsize},
  base/.style={gray!40, line width=0.3pt},
  note/.style={font=\scriptsize, text=navy!85},
  small/.style={font=\scriptsize},
  tarr/.style={{Stealth[length=3pt]}-{Stealth[length=3pt]}, line width=0.5pt, navy!85},
  tguide/.style={navy!35, line width=0.3pt, densely dotted},
]

\def\yR{5.0}\def\yA{4.0}\def\yB{3.15}\def\yC{2.45}\def\yD{1.55}\def\yRx{0.4}
\def\xL{0.5}\def\xR{18.3}

\begin{scope}[on background layer]
\fill[okband]    (1.9,-0.15)  rectangle (5.8,4.45);   
\fill[emptyband] (7.5,-0.15)  rectangle (8.5,4.45);   
\fill[collband]  (10.2,-0.15) rectangle (12.6,4.45);  
\fill[okband]    (14.65,-0.15) rectangle (18.05,4.45);
\end{scope}

\draw[decorate,decoration={brace,amplitude=5pt},navy!70]
 (\xL,6.05) -- (18.05,6.05) node[midway,above=4pt,font=\scriptsize\bfseries,text=navy!80]{One inventory round (frame of $2^{Q}$ slots)};

\foreach \x/\k/\out/\col in {3.85/0/singleton/greenlab, 8.0/1/empty/graylab, 11.4/2/collision/redlab, 16.35/3/singleton/greenlab}{
  \node[small,text=\col] at (\x,5.8){Slot \k};
  \node[font=\scriptsize\bfseries,text=\col] at (\x,5.5){\out};
}

\node[row] at (\xL-0.12,\yR){Reader};
\node[row] at (\xL-0.12,\yA){Tag A};
\node[row] at (\xL-0.12,\yB){Tag B};
\node[row] at (\xL-0.12,\yC){Tag C};
\node[row] at (\xL-0.12,\yD){Tag D};
\node[row,font=\scriptsize\bfseries] at (\xL-0.12,\yRx){Received};
\foreach \y in {\yA,\yB,\yC,\yD,\yRx} \draw[base] (\xL,\y)--(\xR,\y);

\fill[cwc, draw=navy!30] (\xL,\yR-0.27) rectangle (\xR,\yR+0.27);
\node[cmd, minimum width=1.0cm]  (Q)   at (1.0,\yR){Query};
\node[cmd, minimum width=0.7cm]  (ACK0)at (3.9,\yR){ACK};
\node[cmd, minimum width=1.5cm]        at (6.6,\yR){QueryRep};
\node[cmd, minimum width=1.5cm]        at (9.3,\yR){QueryRep};
\node[cmd, minimum width=1.5cm]        at (13.7,\yR){QueryRep};
\node[cmd, minimum width=0.7cm]        at (16.4,\yR){ACK};
\node[small,text=figred!70!black] at (11.4,\yR-0.42){no ACK sent};

\fm{2.15}{\yA-0.16}{figgreen!80!black}{0.32}
\node[small,text=figgreen!55!black,anchor=south] at (2.6,\yA+0.14){RN16};
\fm{4.75}{\yA-0.16}{figgreen!80!black}{0.32}
\node[small,text=figgreen!55!black,anchor=south] at (5.2,\yA+0.14){EPC};
\fm{10.9}{\yB-0.16}{figblue}{0.3}
\node[small,text=figblue,anchor=south] at (11.35,\yB+0.14){RN16};
\fm{11.05}{\yC-0.16}{figred}{0.3}
\node[small,text=figred,anchor=south] at (11.5,\yC+0.14){RN16};
\node[font=\scriptsize,text=figred!75!black,anchor=north,align=center] at (11.4,\yC-0.42){Tags B and C chose the same slot};
\fm{14.9}{\yD-0.16}{figorange}{0.3}
\node[small,text=figorange!80!black,anchor=south] at (15.35,\yD+0.14){RN16};
\fm{17.1}{\yD-0.16}{figorange}{0.3}
\node[small,text=figorange!80!black,anchor=south] at (17.55,\yD+0.14){EPC};

\coordinate (tly) at (0,\yA+0.55);
\draw[tguide] (Q.south east) -- (Q.south east |- tly);
\draw[tguide] (2.15,\yA+0.16) -- (2.15,\yA+0.55);
\draw[tarr] (Q.south east |- tly) -- (2.15,\yA+0.55) node[midway,above=-1pt,small,text=navy!85]{$T_1$};
\draw[tguide] (3.05,\yA+0.16) -- (3.05,\yA+0.55);
\draw[tguide] (ACK0.south west) -- (ACK0.south west |- tly);
\draw[tarr] (3.05,\yA+0.55) -- (ACK0.south west |- tly) node[midway,above=-1pt,small,text=navy!85]{$T_2$};

\fm{2.15}{\yRx-0.16}{figgreen!75!black}{0.34}
\node[small,text=figgreen!55!black,anchor=north] at (2.6,\yRx-0.22){decodable};
\draw[gray!55,line width=0.5pt] (7.7,\yRx) sin (7.9,\yRx+0.07) cos (8.1,\yRx) sin (8.3,\yRx-0.06) cos (8.5,\yRx);
\node[small,text=gray!65,anchor=north] at (8.0,\yRx-0.22){no response};
\draw[navy,line width=1.05pt]
 (10.9,\yRx-0.16)-- ++(0.15,0)-- ++(0,0.54)-- ++(0.12,0)-- ++(0,-0.32)-- ++(0.13,0)-- ++(0,0.44)
 -- ++(0.13,0)-- ++(0,-0.6)-- ++(0.13,0)-- ++(0,0.5)-- ++(0.13,0)-- ++(0,-0.32)-- ++(0.15,0)-- ++(0,0.52)-- ++(0.13,0)-- ++(0,-0.48)-- ++(0.15,0);
\node[small,text=figred!75!black,anchor=north] at (11.55,\yRx-0.22){collision (undecodable)};
\fm{14.9}{\yRx-0.16}{figgreen!75!black}{0.34}
\node[small,text=figgreen!55!black,anchor=north] at (15.35,\yRx-0.22){decodable};

\draw[-{Stealth[length=4pt]}, gray!70, line width=0.5pt] (\xL,-0.55) -- (\xR,-0.55)
  node[anchor=west,font=\scriptsize,text=gray!70]{time};

\begin{scope}[shift={(0.5,-1.2)}]
\fill[cwc,draw=navy!30] (0.1,-0.16) rectangle (0.7,0.16);
\node[anchor=west,font=\scriptsize] at (0.8,0){continuous wave (CW)};
\node[cmd,minimum width=0.55cm] at (4.2,0){}; \node[anchor=west,font=\scriptsize] at (4.55,0){reader command};
\fm{6.9}{-0.16}{figgreen!80!black}{0.28}\node[anchor=west,font=\scriptsize] at (8.05,0){tag backscatter (FM0)};
\draw[navy,line width=1.05pt](11.0,-0.16)--++(0.12,0)--++(0,0.32)--++(0.1,0)--++(0,-0.32)--++(0.12,0)--++(0,0.24)--++(0.1,0)--++(0,-0.24)--++(0.12,0);
\node[anchor=west,font=\scriptsize] at (11.8,0){received collision (sum)};
\end{scope}

\end{tikzpicture}}
\caption{EPC Gen2 inventory round illustrating the reader commands, tag responses, and received signals across empty, singleton, and collision slots. The intervals $T_1$ and $T_2$ denote the time from a reader command to a tag response and from a tag response to the next reader command, respectively.}
\label{fig:protocol}
\vspace{-0.5cm}
\end{figure*}

Collisions among tag responses are a major limitation of passive UHF-RFID. When two or more tags respond in the same slot, their backscattered signals overlap, and a conventional reader fails to decode the collided responses. To reduce such contention, the EPC Gen2 standard adopts framed slotted ALOHA (FSA), where each tag randomly selects a frame slot \cite{global2008epc}. Despite its simplicity, conventional FSA achieves a maximum throughput of only $1/e \approx 0.368$ tags per slot when the frame length approximately equals the tag population \cite{schoute1983dynamic}. A collided slot, however, still contains recoverable information about the responding tags. Recovering the individual tag responses from the collided signal at the physical layer would convert these wasted slots into successful reads and improve the achievable throughput well beyond the FSA bound.

At the medium access control (MAC) layer, anti-collision protocols reduce the probability of collisions rather than recover the collided responses. Dynamic framed slotted ALOHA (DFSA) extends FSA by adapting the frame length to an estimate of the active tag population \cite{vogt2002efficient}. Tag population estimation has been studied using Bayesian inference \cite{floerkemeier2006bayesian}, slot-wise maximum likelihood \cite{knerr2010slot}, multi-frame estimators \cite{vales2011multiframe}, and unified frameworks that combine multiple estimators \cite{chen2014unified}. However, these protocols still treat each collision as undecodable, limiting their throughput to $1/e$. 

Physical layer collision recovery extracts the individual tag responses from the combined signal instead of discarding it. Early receivers based on signal constellation analysis and parameter estimation jointly decode two tags \cite{shen2009separation,khasgiwale2009extracting}, while antenna arrays combined with blind source separation resolve more than two overlapping responses \cite{yu2008anti,mindikoglu2008separation}. The framework in \cite{angerer2010single,angerer2010rfid} estimates channel coefficients and applies zero forcing or successive interference cancellation to recover two colliding tags. Subsequent extensions include multi-antenna processing \cite{kaitovic2011rfid,kaitovic2012channel,kaitovic2013smart}, multipacket reception frameworks for backscatter \cite{fyhn2011multipacket,kang2011decoding}, and coherent reception with software-defined radios (SDRs) \cite{bletsas2012single,kargas2015fullycoherent}.

Other receiver designs reduce reliance on explicit channel estimation or use alternative tag separation methods. These include voltage clustering \cite{tan2016collision}, coding and parallel decoding \cite{mahdavifar2015coding,ou2015come}, and parallel backscatter systems such as FlipTracer and Hubble, which demonstrate concurrent WISP decoding and modified MAC behavior \cite{jin2019fliptracer,jin2018parallelbackscatter}. Blind and statistical methods avoiding pilot-based channel estimation include semi-blind cumulant separation \cite{zhang2020semi}, finite-symbol Viterbi decoding \cite{wu2021fsvd}, and Matrix Pencil parameter estimation \cite{salah2024matrixpencil}. 

Recent receiver designs include mReader \cite{pirayesh2023mreader}, a multi-user multiple-input multiple-output (MU-MIMO) beamforming receiver demonstrated with two tags, a fast Monte Carlo clustering procedure for in-phase and quadrature (I/Q) signal separation \cite{zeng2024fastmc}, and a pilot-free multi-antenna formulation based on the zero constant modulus criterion \cite{siala2025antennaarray}. The geometric approach in \cite{alfayoumi2025nextgen} reports recovery of up to four tags through I/Q clustering and alignment with the ideal constellation. 

While prior collision recovery methods have improved the utilization of collided slots, many first convert the received waveform into an explicit intermediate estimate, e.g., channel state information (CSI), separated constellation clusters, or signal parameters. Estimation errors can propagate to the decoded bits, especially when overlapping responses produce ambiguous I/Q patterns. Deep learning has recently been used for wireless physical layer detection, where end-to-end models can learn a direct mapping from received samples to decoded data instead of estimating intermediate quantities separately \cite{oshea2017introduction,ye2018power,samuel2019learning}. This capability is particularly useful for multi-tag collision recovery in UHF-RFID systems, where decoding can exploit both the I/Q geometry of the superposed responses and the temporal constraints imposed by FM0 encoding. Accordingly, we develop an end-to-end decoding method based on deep learning that recovers individual tag responses directly from the received baseband samples without requiring tag modifications, additional pilot symbols, or CSI. 

In a preliminary version of this work \cite{akyildiz2022mlaided}, neural networks were used to estimate the tag count from the collided signal and the channel coefficients from four additional pilot symbols, followed by maximum likelihood detection. This paper significantly extends that work by replacing the multi-stage pipeline with an end-to-end decoding algorithm and by validating the resulting decoder on measured backscatter signals. Specifically, we propose Self-Attention Tag Recovery (SATR), which builds on a transformer architecture and jointly estimates the number of responding tags and their 16-bit random number (RN16) sequences from the collided waveform.

Table~\ref{tab:comparison} summarizes the main distinctions between SATR and representative prior methods. The main contributions are summarized as follows:
\begin{enumerate}[label=\arabic*)]
\item We formulate multi-tag collision recovery as a set prediction problem and propose SATR, an end-to-end transformer decoding algorithm that maps the baseband I/Q samples of a collided slot directly to an unordered set of RN16 sequences.
\item We design the SATR architecture with FM0-aware input tokens, a self-attention waveform encoder, learned candidate tag representations, and dedicated output modules for tag presence detection, auxiliary channel estimation, and sequential bit decoding. Optimal assignment makes the training objective invariant to candidate ordering, while the fixed candidate set allows a single trained model to handle all considered collision sizes.
\item We numerically evaluate SATR across collision sizes against the ideal FSA limits derived for different recovery configurations. Without requiring CSI, SATR approaches the performance of the Bahl--Cocke--Jelinek--Raviv (BCJR) detector with perfect CSI \cite{bahl1974optimal} and achieves higher throughput than the I/Q geometry method \cite{alfayoumi2025nextgen}. It reaches $0.815$ and $1.87$ tags per slot under single acknowledgment and full recovery, corresponding to $2.2$ and $5.1$ times the conventional FSA limit, respectively.
\item We validate SATR using backscatter measurements from commercial UHF-RFID tags collected with a bistatic SDR reader. SATR correctly estimates the collision size in all direct captures involving up to four simultaneous responses, while composite collisions with known transmitted bits demonstrate high tag success rates consistent with the numerical results.
\end{enumerate}

The remainder of this paper is organized as follows. Section~\ref{sec:system_model} presents the system and signal model and analyzes the FSA throughput under multi-tag collision recovery. Section~\ref{sec:proposed_approach} introduces SATR and describes its self-attention architecture, training, and inference procedures. Section~\ref{sec:results} presents the numerical evaluation of SATR under different collision and recovery settings. Section~\ref{sec:experimental_results} validates SATR using measured backscatter signals. Section~\ref{sec:conclusion} concludes the paper.

\section{System Model}
\label{sec:system_model}
This section represents the system model for a passive UHF-RFID inventory system operating under the EPC Gen2 standard \cite{global2008epc}. We describe the inventory procedure and the slot states that produce tag collisions. We then analyze the FSA throughput under conventional decoding and collision recovery. Finally, we derive the baseband model for a collided slot and describe the resulting I/Q constellation structure.

\subsection{Notation Convention}

Throughout this paper, vectors and matrices are denoted by boldface lowercase and uppercase letters, e.g., $\mathbf{x}$ and $\mathbf{X}$, respectively, and $(\cdot)^{\mathsf{T}}$ denotes the transpose. The sets $\R$ and $\C$ denote the real and complex numbers. The operator $\E[\cdot]$ denotes expectation, while $\operatorname{Re}\{\cdot\}$ and $\operatorname{Im}\{\cdot\}$ denote the real and imaginary parts of a complex quantity. For a complex scalar $x$, $|x|$ and $\angle x$ denote its magnitude and phase, respectively, and $\|\mathbf{x}\|$ denotes the Euclidean norm of vector $\mathbf{x}$. The identity matrix of size $n$ is denoted by $\mathbf{I}_n$, $\mathbf{1}_n$ denotes the all-ones vector of length $n$, $\CN(\boldsymbol{\mu},\boldsymbol{\Sigma})$ denotes the circularly symmetric complex Gaussian (CSCG) distribution with mean $\boldsymbol{\mu}$ and covariance $\boldsymbol{\Sigma}$, and $\mathcal{U}(a,b)$ denotes the uniform distribution over $(a,b)$. The operator $\oplus$ denotes modulo two addition, and the indicator function is denoted by $\mathbbm{1}\{\cdot\}$.

\subsection{EPC Gen2 Protocol}
\label{subsec:epc_gen2}

The EPC Gen2 standard defines a half-duplex, reader-initiated protocol for passive UHF-RFID communication \cite{global2008epc}. During an inventory round, the reader sends commands by modulating its RF carrier and continues transmitting a CW during tag responses. This signal powers the tags and provides the carrier for their backscatter responses. Each tag responds by switching its antenna impedance between two load states, thereby modulating the signal reflected toward the reader.

Fig.~\ref{fig:protocol} shows the EPC Gen2 inventory sequence. The reader begins an inventory round by sending a Query command with the frame size parameter $Q$, which determines $K=2^Q$ slots. After receiving the Query command, each tag independently and uniformly selects a slot counter value from $\{0,1,\ldots,K-1\}$. A tag responds only when its counter is zero. The reader advances the inventory frame using QueryRep commands, where each QueryRep decrements the slot counter of every tag that has not responded yet.

A tag whose slot counter reaches zero backscatters an RN16 using FM0 encoding. If the reader decodes the RN16, it sends an acknowledgment (ACK) containing the same RN16 value. After verifying the ACK, the tag backscatters its Electronic Product Code (EPC), completing the identification exchange. The timing is governed by the intervals $T_1$ and $T_2$, where $T_1$ is the time from the end of a reader command to the start of the tag response and $T_2$ is the time from the end of the tag response to the next reader command.

The inventory procedure produces three slot states: i) An empty slot: no tag responds. ii) A singleton slot: one tag responds. iii) A collision slot: two or more tags respond. Under conventional EPC Gen2 decoding, only a singleton slot leads to the ACK and EPC exchange. A collision slot is discarded because the overlapping RN16 waveforms are not decoded. This slot structure underlies the FSA throughput limit and motivates physical layer recovery of collided responses.

\subsection{Framed Slotted ALOHA and Throughput Analysis}
\label{subsec:fsa}
The FSA throughput is determined by the distribution of tag responses across the slots of a frame. We model an FSA frame with $N$ participating tags and $K$ slots. The frame size controls the balance between empty and collision slots. A smaller $K$ increases the probability of collisions because fewer slots are available, whereas a larger $K$ reduces collisions at the cost of more empty slots.

Under the standard FSA model, each tag independently and uniformly selects one of the $K$ slots. Hence, the number of tag responses in any given slot, denoted by $R$, follows a binomial distribution. The probability that a slot contains exactly $R$ tag responses is
\begin{equation}
	P(R) = \binom{N}{R} \left( \frac{1}{K} \right)^R \left( 1 - \frac{1}{K} \right)^{N-R}.
	\label{eq:slot_prob}
\end{equation}
Let $\mathcal{X}_R$ denote the number of slots in the frame that contain exactly $R$ tag responses. Since each slot has probability $P(R)$ of containing $R$ responses, the expected number of such slots is $\E[\mathcal{X}_R]=KP(R)$. The conventional FSA throughput $\eta_{\text{conv}}$ is defined as the expected number of acknowledged tags per slot. Because a conventional EPC Gen2 reader decodes only singleton slots, the expected throughput is
\begin{equation}
	\eta_{\text{conv}} = \frac{\E[\mathcal{X}_1]}{K} = \frac{N}{K} \left( 1 - \frac{1}{K} \right)^{N-1}.
	\label{eq:throughput_conv}
\end{equation}
For large $N$ and $K$ with fixed load $\rho = N/K$, the binomial distribution in \eqref{eq:slot_prob} is well approximated by a Poisson distribution with mean $\rho$, so that $P(R) \approx e^{-\rho}\frac{\rho^R}{R!}$ and the throughput depends on $N$ and $K$ only through $\rho$ \cite{schoute1983dynamic}. The conventional throughput then becomes $\eta_{\text{conv}} \approx \rho e^{-\rho}$, which attains its maximum $\eta_{\text{conv}}^* = 1/e \approx 0.368$ at $\rho = 1$.

Collision recovery modifies the FSA throughput model by allowing collided slots to contribute decoded tags. Let $M$ denote the maximum collision size that can be resolved, and let $J$ denote the maximum number of decoded tag responses that can be acknowledged from a recoverable slot, with $1\leq J\leq M$. A slot containing $R$ tag responses is recoverable if $R\leq M$, and remains unresolved otherwise. From a recoverable slot, the reader decodes the collided responses and acknowledges up to $J$ of them, corresponding to $\min(R,J)$ acknowledged tags. The resulting expected throughput is
\begin{align}
	\eta_{M,J}
	&= \frac{1}{K} \sum_{R=1}^{M} \min(R,J)\, \E[\mathcal{X}_R]
	= \sum_{R=1}^{M} \min(R,J)\, P(R) \nonumber \\
	&\approx e^{-\rho} \sum_{R=1}^{M} \min(R,J)\, \frac{\rho^R}{R!}.
	\label{eq:throughput_general}
\end{align}
The case $J=1$ represents the standard EPC Gen2 acknowledgment constraint, under which the reader can acknowledge at most one decoded tag response. Setting $J=1$ in \eqref{eq:throughput_general} gives
\begin{equation}
	\eta_{M,1}
	= \frac{1}{K} \sum_{R=1}^{M} \E[\mathcal{X}_R]
	= \sum_{R=1}^{M} P(R)
	\approx e^{-\rho} \sum_{R=1}^{M} \frac{\rho^R}{R!}.
	\label{eq:throughput_j1}
\end{equation}

\begin{figure}[h]
\centering
\begin{tikzpicture}
\begin{axis}[
	width=\columnwidth, height=0.6\columnwidth,
	xlabel={Frame-to-tag ratio $K/N$}, ylabel={Expected throughput $\eta_{M,1}$},
	xmin=0, xmax=2, ymin=0, ymax=1.08, ytick={0,0.2,0.4,0.6,0.8},
	grid=both, grid style={line width=0.08pt, draw=gray!15},
	tick label style={font=\scriptsize}, label style={font=\footnotesize}, clip=false,
	legend style={at={(0.5,-0.26)}, anchor=north, legend columns=4, font=\scriptsize, draw=none, /tikz/every even column/.append style={column sep=6pt}}
]
\addplot[very thick, figcolor2, mark=none, domain=0.1:2, samples=200] {exp(-1/x)*(1/x)};\addlegendentry{$M{=}1$}
\addplot[very thick, figcolor5, mark=none, domain=0.1:2, samples=200] {exp(-1/x)*((1/x)+(1/x)^2/2)};\addlegendentry{$M{=}2$}
\addplot[very thick, figcolor3, mark=none, domain=0.1:2, samples=200] {exp(-1/x)*((1/x)+(1/x)^2/2+(1/x)^3/6)};\addlegendentry{$M{=}3$}
\addplot[very thick, figcolor1, mark=none, domain=0.1:2, samples=200] {exp(-1/x)*((1/x)+(1/x)^2/2+(1/x)^3/6+(1/x)^4/24)};\addlegendentry{$M{=}4$}
\addplot[only marks, mark=*, mark size=1.6pt, color=black, forget plot] coordinates {(1.0,0.368)(0.707,0.587)(0.55,0.726)(0.452,0.817)};
\draw[black,line width=0.5pt](axis cs:1.15,1.08)--(axis cs:2.0,1.08);
\node[font=\tiny,anchor=base]at(axis cs:1.27,1.035){$M$};
\node[font=\tiny,anchor=base]at(axis cs:1.55,1.035){$(K/N)^*$};
\node[font=\tiny,anchor=base]at(axis cs:1.86,1.035){$\eta_{M,1}^*$};
\draw[black,line width=0.3pt](axis cs:1.15,0.995)--(axis cs:2.0,0.995);
\node[font=\tiny,anchor=base]at(axis cs:1.27,0.91){\textcolor{figcolor1}{$4$}};\node[font=\tiny,anchor=base]at(axis cs:1.55,0.91){\textcolor{figcolor1}{$0.452$}};\node[font=\tiny,anchor=base]at(axis cs:1.86,0.91){\textcolor{figcolor1}{$0.817$}};
\node[font=\tiny,anchor=base]at(axis cs:1.27,0.82){\textcolor{figcolor3}{$3$}};\node[font=\tiny,anchor=base]at(axis cs:1.55,0.82){\textcolor{figcolor3}{$0.550$}};\node[font=\tiny,anchor=base]at(axis cs:1.86,0.82){\textcolor{figcolor3}{$0.726$}};
\node[font=\tiny,anchor=base]at(axis cs:1.27,0.73){\textcolor{figcolor5}{$2$}};\node[font=\tiny,anchor=base]at(axis cs:1.55,0.73){\textcolor{figcolor5}{$0.707$}};\node[font=\tiny,anchor=base]at(axis cs:1.86,0.73){\textcolor{figcolor5}{$0.587$}};
\node[font=\tiny,anchor=base]at(axis cs:1.27,0.64){\textcolor{figcolor2}{$1$}};\node[font=\tiny,anchor=base]at(axis cs:1.55,0.64){\textcolor{figcolor2}{$1.000$}};\node[font=\tiny,anchor=base]at(axis cs:1.86,0.64){\textcolor{figcolor2}{$0.368$}};
\draw[black,line width=0.5pt](axis cs:1.15,0.61)--(axis cs:2.0,0.61);
\draw[optarrow](axis cs:0.452,0.817)--(axis cs:1.15,0.91);
\draw[optarrow](axis cs:0.55,0.726)--(axis cs:1.15,0.82);
\draw[optarrow](axis cs:0.707,0.587)--(axis cs:1.15,0.73);
\draw[optarrow](axis cs:1.0,0.368)--(axis cs:1.15,0.64);
\end{axis}
\end{tikzpicture}
\vspace{-0.1cm}
\caption{FSA throughput under the single acknowledgment setting ($J=1$) versus the frame-to-tag ratio $K/N$ for different values of $M$.}
\label{fig:throughput_kn}
\vspace{-0.4cm}
\end{figure}
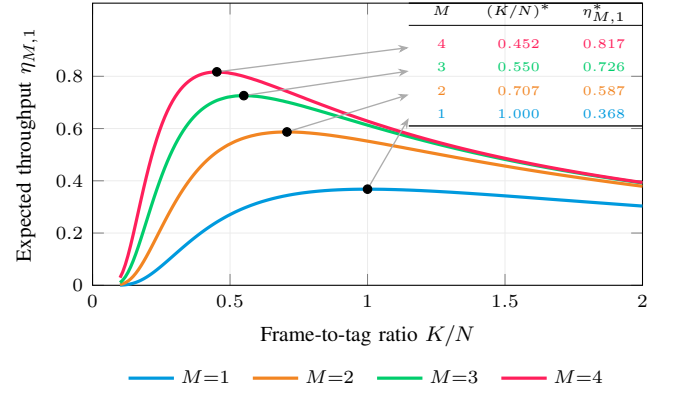

In the single acknowledgment setting, $\eta_{M,1}$ denotes the expected throughput when collisions of up to $M$ tags are recoverable. Fig.~\ref{fig:throughput_kn} shows $\eta_{M,1}$ versus the frame-to-tag ratio $K/N$ for $M \in \{1,2,3,4\}$, where $M=1$ reduces to conventional FSA.\footnote{We consider recovery configurations up to $M=4$ since the probability of a slot containing $R$ responses decays rapidly with $R$ under the slot distribution in \eqref{eq:slot_prob}. Collisions of more than four tags are therefore rare at practical frame loads and would contribute little additional throughput even if recoverable.} The inset table reports the optimal frame-to-tag ratio and maximum throughput for each curve. Increasing $M$ makes larger collisions recoverable, which increases the maximum throughput and shifts the optimal $K/N$ below one. For $M=4$, the maximum throughput is $0.817$ at $K/N=0.452$, corresponding to a $2.22$-fold gain over conventional FSA.

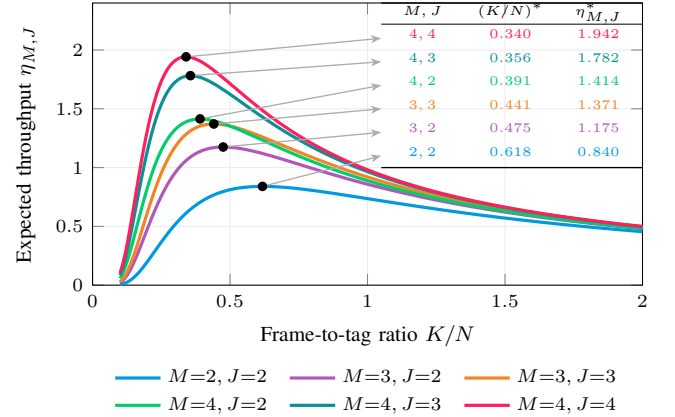
\begin{figure}[t]
\centering
\begin{tikzpicture}
\begin{axis}[
	width=\columnwidth, height=0.6\columnwidth,
	xlabel={Frame-to-tag ratio $K/N$}, ylabel={Expected throughput $\eta_{M,J}$},
	xmin=0, xmax=2, ymin=0, ymax=2.4, ytick={0,0.5,1,1.5,2},
	grid=both, grid style={line width=0.08pt, draw=gray!15},
	tick label style={font=\scriptsize}, label style={font=\footnotesize}, clip=false,
	legend style={at={(0.5,-0.26)}, anchor=north, legend columns=3, font=\scriptsize, draw=none, /tikz/every even column/.append style={column sep=6pt}}
]
\addplot[very thick, figcolor2, mark=none, domain=0.1:2, samples=200] {exp(-1/x)*((1/x)+(1/x)^2)};\addlegendentry{$M{=}2,J{=}2$}
\addplot[very thick, figcolor4, mark=none, domain=0.1:2, samples=200] {exp(-1/x)*((1/x)+(1/x)^2+(1/x)^3/3)};\addlegendentry{$M{=}3,J{=}2$}
\addplot[very thick, figcolor5, mark=none, domain=0.1:2, samples=200] {exp(-1/x)*((1/x)+(1/x)^2+(1/x)^3/2)};\addlegendentry{$M{=}3,J{=}3$}
\addplot[very thick, figcolor3, mark=none, domain=0.1:2, samples=200] {exp(-1/x)*((1/x)+(1/x)^2+(1/x)^3/3+(1/x)^4/12)};\addlegendentry{$M{=}4,J{=}2$}
\addplot[very thick, figcolor6, mark=none, domain=0.1:2, samples=200] {exp(-1/x)*((1/x)+(1/x)^2+(1/x)^3/2+(1/x)^4/8)};\addlegendentry{$M{=}4,J{=}3$}
\addplot[very thick, figcolor1, mark=none, domain=0.1:2, samples=200] {exp(-1/x)*((1/x)+(1/x)^2+(1/x)^3/2+(1/x)^4/6)};\addlegendentry{$M{=}4,J{=}4$}
\addplot[only marks, mark=*, mark size=1.6pt, color=black, forget plot] coordinates {(0.618,0.840)(0.475,1.175)(0.441,1.371)(0.391,1.414)(0.356,1.782)(0.340,1.942)};
\draw[black,line width=0.5pt](axis cs:1.05,2.4)--(axis cs:2.0,2.4);
\node[font=\tiny,anchor=base]at(axis cs:1.2,2.3){$M,J$};
\node[font=\tiny,anchor=base]at(axis cs:1.52,2.3){$(K/N)^*$};
\node[font=\tiny,anchor=base]at(axis cs:1.84,2.3){$\eta_{M,J}^*$};
\draw[black,line width=0.3pt](axis cs:1.05,2.22)--(axis cs:2.0,2.22);
\node[font=\tiny,anchor=base]at(axis cs:1.2,2.10){\textcolor{figcolor1}{$4,4$}};\node[font=\tiny,anchor=base]at(axis cs:1.52,2.10){\textcolor{figcolor1}{$0.340$}};\node[font=\tiny,anchor=base]at(axis cs:1.84,2.10){\textcolor{figcolor1}{$1.942$}};
\node[font=\tiny,anchor=base]at(axis cs:1.2,1.90){\textcolor{figcolor6}{$4,3$}};\node[font=\tiny,anchor=base]at(axis cs:1.52,1.90){\textcolor{figcolor6}{$0.356$}};\node[font=\tiny,anchor=base]at(axis cs:1.84,1.90){\textcolor{figcolor6}{$1.782$}};
\node[font=\tiny,anchor=base]at(axis cs:1.2,1.70){\textcolor{figcolor3}{$4,2$}};\node[font=\tiny,anchor=base]at(axis cs:1.52,1.70){\textcolor{figcolor3}{$0.391$}};\node[font=\tiny,anchor=base]at(axis cs:1.84,1.70){\textcolor{figcolor3}{$1.414$}};
\node[font=\tiny,anchor=base]at(axis cs:1.2,1.50){\textcolor{figcolor5}{$3,3$}};\node[font=\tiny,anchor=base]at(axis cs:1.52,1.50){\textcolor{figcolor5}{$0.441$}};\node[font=\tiny,anchor=base]at(axis cs:1.84,1.50){\textcolor{figcolor5}{$1.371$}};
\node[font=\tiny,anchor=base]at(axis cs:1.2,1.30){\textcolor{figcolor4}{$3,2$}};\node[font=\tiny,anchor=base]at(axis cs:1.52,1.30){\textcolor{figcolor4}{$0.475$}};\node[font=\tiny,anchor=base]at(axis cs:1.84,1.30){\textcolor{figcolor4}{$1.175$}};
\node[font=\tiny,anchor=base]at(axis cs:1.2,1.10){\textcolor{figcolor2}{$2,2$}};\node[font=\tiny,anchor=base]at(axis cs:1.52,1.10){\textcolor{figcolor2}{$0.618$}};\node[font=\tiny,anchor=base]at(axis cs:1.84,1.10){\textcolor{figcolor2}{$0.840$}};
\draw[black,line width=0.5pt](axis cs:1.05,1.0)--(axis cs:2.0,1.0);
\draw[optarrow](axis cs:0.340,1.942)--(axis cs:1.05,2.10);
\draw[optarrow](axis cs:0.356,1.782)--(axis cs:1.05,1.90);
\draw[optarrow](axis cs:0.391,1.414)--(axis cs:1.05,1.70);
\draw[optarrow](axis cs:0.441,1.371)--(axis cs:1.05,1.50);
\draw[optarrow](axis cs:0.475,1.175)--(axis cs:1.05,1.30);
\draw[optarrow](axis cs:0.618,0.840)--(axis cs:1.05,1.10);
\end{axis}
\end{tikzpicture}
\vspace{-0.1cm}
\caption{FSA throughput under multi-tag recovery ($J>1$) versus the frame-to-tag ratio $K/N$ for different recovery configurations $(M,J)$.}
\label{fig:throughput_mj}
\vspace{-0.5cm}
\end{figure}

Multi-tag recovery corresponds to $J>1$ and further improves throughput by allowing more than one decoded tag to be acknowledged. As $J$ approaches $M$, the model moves toward full recovery, where every decoded response from any collision of size at most $M$ can be acknowledged. Fig.~\ref{fig:throughput_mj} depicts $\eta_{M,J}$ for representative $(M,J)$ pairs with $J>1$, and the inset table lists the optimal operating points. The curves show that increasing either $M$ or $J$ raises the maximum throughput and shifts the optimal $K/N$ ratio downward, because recoverable collisions make denser frames more efficient. For $(M,J)=(4,4)$, the throughput reaches $\eta_{4,4}^* = 1.942$ tags per slot, more than five times the conventional FSA limit.

\subsection{Communication Model}
\label{subsec:comm_model}

\begin{figure}[t]
\centering
\begin{tikzpicture}[>=Latex, line cap=round, font=\footnotesize,
  fwd/.style={fwdcol, line width=1.0pt, ->},
  bwd/.style={bwdcol, line width=1.0pt, dash pattern=on 3pt off 2pt, ->},
  lbl/.style={fill=white, inner sep=1pt, font=\footnotesize}]

  \node[draw=black, line width=0.9pt, rounded corners=2pt, fill=readercol!85,
        fill opacity=0.72, minimum width=1.75cm, minimum height=2.0cm] (rd) at (0,0) {};
  \node[font=\bfseries] at (0,0.66) {Reader};
  \node[draw=fwdcol, line width=0.9pt, fill=white, minimum width=1.15cm,
        minimum height=0.38cm] (tx) at (0,0.16) {\scriptsize TX};
  \node[draw=bwdcol, line width=0.9pt, fill=white, minimum width=1.15cm,
        minimum height=0.38cm] (rx) at (0,-0.40) {\scriptsize RX};

  \draw[line width=1.0pt] (0.875,0) -- (1.15,0);
  \fill[black!55] (1.15,-0.27) -- (1.15,0.27) -- (1.52,0) -- cycle;
  \coordinate (H) at (1.52,0);

  \draw[fwdcol, line width=1.0pt, decorate,
        decoration={snake, amplitude=1.5pt, segment length=5pt}]
        (-0.45,1.28) -- (0.15,1.28);
  \node[fwdcol, anchor=west, font=\scriptsize] at (0.18,1.28) {CW};

  \draw[leakcol, dash pattern=on 3pt off 2pt, line width=0.9pt, ->]
        (tx.west) .. controls (-1.65,0.16) and (-1.65,-0.40) .. (rx.west);
  \node[leakcol, fill=white, inner sep=0.6pt] at (-1.38,-0.12) {$\mathbf{L}$};

  \def\tagx{5.5}
  \tikzset{tagbox/.style={draw=black, line width=0.9pt, rounded corners=2pt,
        fill=tagcol!80, fill opacity=0.72, text opacity=1,
        minimum width=1.55cm, minimum height=0.86cm, align=center}}
  \node[tagbox] (t1) at (\tagx, 1.80) {\textbf{Tag 1}\\[-2pt]$a_1(t)$};
  \node[tagbox] (t2) at (\tagx, 0.0) {\textbf{Tag 2}\\[-2pt]$a_2(t)$};
  \node at (\tagx,-0.90) {$\vdots$};
  \node[tagbox] (tR) at (\tagx,-1.80) {\textbf{Tag $R$}\\[-2pt]$a_R(t)$};

  \draw[fwd] ($(H)+(0,0.05)$) -- ($(t1.west)+(0,0.13)$)
        node[sloped, above, pos=0.72, text=fwdcol, inner sep=2pt]{$h_1^f$};
  \draw[bwd] ($(t1.west)+(0,-0.13)$) -- ($(H)+(0,-0.05)$);
  \path ($(H)+(0,-0.05)$) -- ($(t1.west)+(0,-0.13)$)
        node[sloped, below, pos=0.72, text=bwdcol, inner sep=2pt]{$\mathbf{h}_1^b$};

  \draw[fwd] ($(H)+(0,0.05)$) -- ($(t2.west)+(0,0.13)$)
        node[sloped, above, pos=0.72, text=fwdcol, inner sep=2pt]{$h_2^f$};
  \draw[bwd] ($(t2.west)+(0,-0.13)$) -- ($(H)+(0,-0.05)$);
  \path ($(H)+(0,-0.05)$) -- ($(t2.west)+(0,-0.13)$)
        node[sloped, below, pos=0.72, text=bwdcol, inner sep=2pt]{$\mathbf{h}_2^b$};

  \draw[fwd] ($(H)+(0,0.05)$) -- ($(tR.west)+(0,0.13)$)
        node[sloped, above, pos=0.72, text=fwdcol, inner sep=2pt]{$h_R^f$};
  \draw[bwd] ($(tR.west)+(0,-0.13)$) -- ($(H)+(0,-0.05)$);
  \path ($(H)+(0,-0.05)$) -- ($(tR.west)+(0,-0.13)$)
        node[sloped, below, pos=0.72, text=bwdcol, inner sep=2pt]{$\mathbf{h}_R^b$};

  \begin{scope}[shift={(-1.4,-3.0)}, font=\scriptsize]
    \draw[fwd] (0,0)--(0.55,0);
      \node[right,inner sep=2pt] at (0.55,0){Forward $h_i^f$};
    \draw[bwd] (2.8,0)--(3.35,0);
      \node[right,inner sep=2pt] at (3.35,0){Backward $\mathbf{h}_i^b$};
    \draw[leakcol, dash pattern=on 3pt off 2pt, line width=0.9pt, ->] (5.7,0)--(6.25,0);
      \node[right,inner sep=2pt] at (6.25,0){Leakage $\mathbf{L}$};
  \end{scope}
\end{tikzpicture}
\caption{Passive UHF RFID backscatter model with $R$ responding tags, including the reader CW, tag backscatter paths, and carrier leakage.}
\label{fig:comm_model}
\vspace{-0.6cm}
\end{figure}
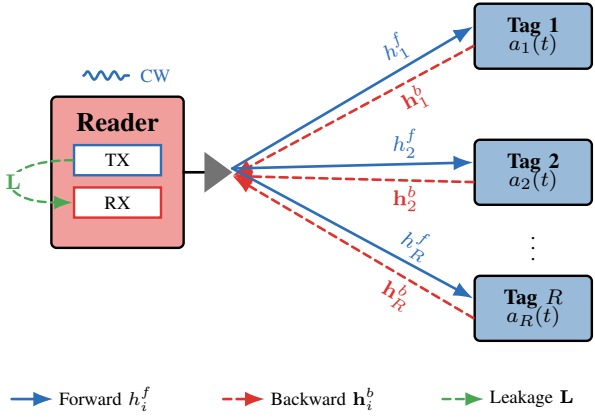

We formulate the physical signal model for a single FSA slot, following the passive UHF-RFID backscatter models in \cite{angerer2010rfid, bletsas2012single}. We consider a reader with $N_{\mathrm{rx}}$ receive antennas and a slot in which $R$ tags respond. During the response interval, the reader continues transmitting its CW, while each tag switches its antenna load according to its FM0 encoded sequence. The reader therefore observes the sum of the tag backscatter signals, carrier leakage, and receiver noise, as illustrated in Fig.~\ref{fig:comm_model}.

A passive tag modulates the incident CW by changing the impedance presented to its antenna \cite{nikitin2007differential}. The tag uses two impedance states. In the absorbing state, most of the incident power is delivered to the tag circuitry, so the reflected component is weak. In the reflecting state, impedance mismatch increases the backscattered component. We represent these two load states by binary levels $0$ and $1$, corresponding to the absorbing and reflecting states, respectively.

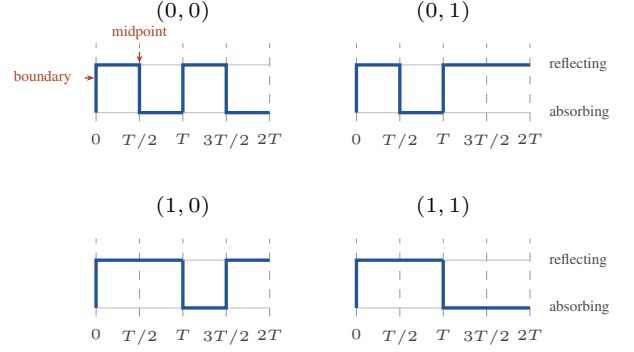
\begin{figure}[!t]
\centering
\resizebox{\columnwidth}{!}{\definecolor{fm0line}{RGB}{35,85,150}
\definecolor{fm0grid}{RGB}{150,150,150}
\definecolor{fm0mark}{RGB}{180,70,40}

\newcommand{\fmpanel}[4]{%
\begin{scope}[shift={(#1,#2)}]
  \def\lo{0}\def\hi{0.55}
  \foreach \gx in {0,0.5,1.0,1.5,2.0}
    \draw[fm0grid,dashed,line width=0.3pt] (\gx,-0.1) -- (\gx,0.8);
  \draw[fm0grid!65,line width=0.25pt] (0,\lo) -- (2.0,\lo);
  \draw[fm0grid!65,line width=0.25pt] (0,\hi) -- (2.0,\hi);
  \ifnum#3=0
    \draw[fm0line,line width=1.1pt] (0,\lo)--(0,\hi)--(0.5,\hi)--(0.5,\lo)--(1.0,\lo);
    \def\enda{\lo}
  \else
    \draw[fm0line,line width=1.1pt] (0,\lo)--(0,\hi)--(1.0,\hi);
    \def\enda{\hi}
  \fi
  \ifdim\enda pt=\lo pt \def\s{\hi}\else\def\s{\lo}\fi
  \ifnum#4=0
    \ifdim\s pt=\hi pt
      \draw[fm0line,line width=1.1pt] (1.0,\enda)--(1.0,\hi)--(1.5,\hi)--(1.5,\lo)--(2.0,\lo);
    \else
      \draw[fm0line,line width=1.1pt] (1.0,\enda)--(1.0,\lo)--(1.5,\lo)--(1.5,\hi)--(2.0,\hi);
    \fi
  \else
    \draw[fm0line,line width=1.1pt] (1.0,\enda)--(1.0,\s)--(2.0,\s);
  \fi
  \foreach \tx/\tl in {0/{$0$},0.5/{$T/2$},1.0/{$T$},1.5/{$3T/2$},2.0/{$2T$}}
    \node[font=\tiny,anchor=north,text=black!75] at (\tx,-0.12){\tl};
  \node[anchor=south,font=\scriptsize] at (1.0,0.92){$(#3,#4)$};
\end{scope}}

\begin{tikzpicture}[font=\footnotesize, x=1cm, y=1cm]
  \fmpanel{0}{2.25}{0}{0}
  \fmpanel{3.0}{2.25}{0}{1}
  \fmpanel{0}{0}{1}{0}
  \fmpanel{3.0}{0}{1}{1}
  \foreach \y in {2.25,0}{
    \node[anchor=west,font=\tiny,text=black!70] at (5.1,\y+0.55){reflecting};
    \node[anchor=west,font=\tiny,text=black!70] at (5.1,\y){absorbing};
  }
  \node[fm0mark,font=\tiny,anchor=east] at (-0.16,2.25+0.4){boundary};
  \draw[fm0mark,-{Stealth[length=2.5pt]},line width=0.45pt] (-0.13,2.25+0.4) -- (-0.02,2.25+0.4);
  \node[fm0mark,font=\tiny,anchor=south] at (0.5,2.25+0.72){midpoint};
  \draw[fm0mark,-{Stealth[length=2.5pt]},line width=0.45pt] (0.5,2.25+0.7) -- (0.5,2.25+0.57);
\end{tikzpicture}}
\caption{FM0 load states for the four two bit input patterns $(b_i[k],\,b_i[k{+}1])$, assuming the initial state is absorbing.}
\label{fig:fm0_coding}
\vspace{-0.6cm}
\end{figure}

Each responding tag sends an RN16 response encoded using FM0 \cite{global2008epc}. For tag $i$, let $b_i[k]\in\{0,1\}$ denote the $k$-th transmitted RN16 bit, and let $\mathbf{b}_i=[b_i[0],\ldots,b_i[15]]$ denote the RN16 sequence. FM0 maps these bits to the two binary load states. Each bit occupies one symbol interval of duration $T$, which is divided into two half symbols. The load state toggles at the beginning of each symbol interval, and the bit value determines whether an additional transition occurs at the half-symbol boundary. If $b_i[k]=0$, the load state toggles at the half-symbol boundary. If $b_i[k]=1$, the load state remains unchanged during the second half symbol. Thus, FM0 represents each bit by the transition pattern within its symbol interval. Fig.~\ref{fig:fm0_coding} shows the resulting load waveforms for the four two bit patterns.

Let $c_i[k]\in\{0,1\}$ denote the load state during the first half of the $k$-th symbol interval. The initial value $c_i[0]$ sets the starting load state. Following the FM0 transition rule, $c_i[k+1] = c_i[k]\oplus b_i[k]$. The corresponding load waveform $a_i(t)$ is given by
\begin{equation}
	a_i(t)=\sum_{k=0}^{15} p_{i,k}(t-kT),
	\label{eq:tag_modulation}
\end{equation}
where $p_{i,k}(t)$ is the FM0 pulse associated with $b_i[k]$, i.e.,
\begingroup
\small
\begin{equation}
p_{i,k}(t)=
\begin{cases}
	c_i[k]\Pi(t/T), & b_i[k]=1,\\[1mm]
	c_i[k]\Pi(2t/T)+(1-c_i[k])\Pi(2t/T-1), & b_i[k]=0,
\end{cases}
\label{eq:fm0_pulse}
\end{equation}
\endgroup
where $\Pi(x)$ is the unit rectangular pulse.

The CW propagates to tag $i$ through the forward channel $h_i^f \in \C$. The differential reflection between the absorbing and reflecting states is denoted by $\sqrt{\Delta\sigma_i}$, where $|\Delta\sigma_i|$ is the normalized differential radar cross section (RCS) of tag $i$ \cite{nikitin2007differential}. The return path from tag $i$ to the reader is represented by the backward channel $\mathbf{h}_i^b \in \C^{N_{\mathrm{rx}}}$, whose $j$-th entry $h_{i,j}^b$ denotes the channel from tag $i$ to receive antenna $j$. At antenna $j$, the passband signal is modeled as
\begin{align}
	y_j^{\text{pb}}(t) &= \sum_{i=1}^{R} |h_i^f|\, \sqrt{|\Delta\sigma_i|}\, |h_{i,j}^b| \, a_i(t) \cos(2\pi f_c t + \phi_{i,j}) \nonumber \\
	&\quad + s_{\text{leak},j}(t) + n_j^{\text{pb}}(t),
	\label{eq:passband_signal}
\end{align}
where $\phi_{i,j} = \angle h_i^f + \angle \sqrt{\Delta\sigma_i} + \angle h_{i,j}^b$ is the total phase shift of the backscattered component, and $n_j^{\text{pb}}(t)$ is passband additive white Gaussian noise. The leakage term is $s_{\text{leak},j}(t)=|L_j|\cos(2\pi f_c t+\phi_j^{\text{leak}})$, where $|L_j|$ and $\phi_j^{\text{leak}}$ are the leakage amplitude and phase at antenna $j$.

After coherent I/Q downconversion and low pass filtering, the received signal at antenna $j$ is represented by the complex baseband signal $y_j(t)$. Since each response is produced by reflecting the reader CW, all responses share the same carrier reference and do not introduce carrier frequency offset. The leakage term also downconverts to the complex DC component $L_j = |L_j| e^{j\phi_j^{\text{leak}}}$. Collecting the baseband signals of the $N_{\mathrm{rx}}$ receive antennas in $\mathbf{y}(t) = [y_1(t), \ldots, y_{N_{\mathrm{rx}}}(t)] \in \C^{N_{\mathrm{rx}}}$ gives
\begin{equation}
\begin{aligned}
	\mathbf{y}(t)
	&= \sum_{i=1}^{R} h_i^f \sqrt{\Delta\sigma_i} \, \mathbf{h}_i^b \, a_i(t) + \mathbf{L} + \mathbf{n}(t) \\
	&= \sum_{i=1}^{R} \mathbf{g}_i a_i(t) + \mathbf{L} + \mathbf{n}(t),
\end{aligned}
	\label{eq:baseband_signal}
\end{equation}
where $\mathbf{g}_i = h_i^f\sqrt{\Delta\sigma_i}\mathbf{h}_i^b \in \C^{N_{\mathrm{rx}}}$ is the effective channel of tag $i$, $\mathbf{L} \in \C^{N_{\mathrm{rx}}}$ is the carrier leakage vector with $j$-th element $L_j$, and $\mathbf{n}(t) \sim \CN(\mathbf{0}, N_0 \mathbf{I}_{N_{\mathrm{rx}}})$. The average receive signal-to-noise ratio (SNR) per antenna is defined as $\mathrm{SNR} = \sum_{i=1}^{R}\E[\|\mathbf{g}_i\|^2]/(RN_{\mathrm{rx}}N_0)$.

\begin{figure*}[!t]
	\centering
	\resizebox{0.8\textwidth}{!}{\input{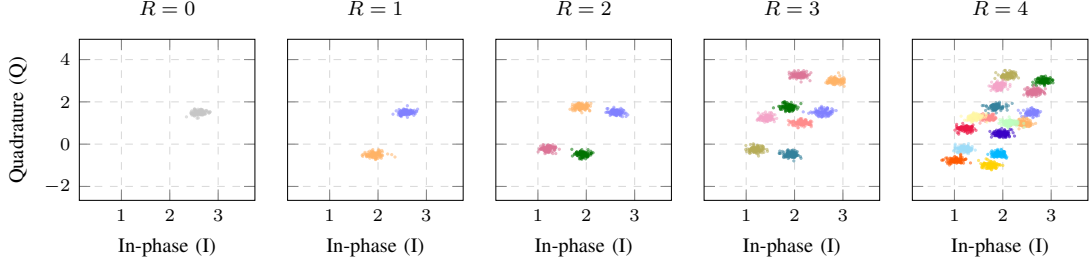}}
	\caption{Synthetic I/Q samples for $R = 0$ to $R = 4$ responding tags at receive antenna $j$. Point colors indicate load-state combinations, and samples are generated at 20~dB SNR.}
	\label{fig:iq_constellation}
	\vspace{-0.6cm}
\end{figure*}

We assume that the RN16 symbol boundaries of the responding tags are aligned within a collision slot. This assumption is consistent with EPC Gen2 operation, where tag responses are reader initiated, occur within the specified turnaround interval after the reader command, and begin with a known FM0 preamble \cite{global2008epc}. It also follows the coherent collision model in \cite{bletsas2012single}, which treats collided responses as having aligned bit boundaries and validates this assumption experimentally. Practical tags can still exhibit small timing and backscatter link frequency variations \cite{angerer2010rfid}, so in the measurement study of Section~\ref{sec:experimental_results}, we locate the response window from the Query timing and extract the measured samples using the nominal RN16 bit grid. This processing provides experimental support for using the synchronized collision model under the considered passive UHF-RFID setup.

\subsection{I/Q Constellation Structure}
\label{subsec:iq_constellation}

The baseband model in \eqref{eq:baseband_signal} characterizes the I/Q constellation observed during a collision. For receive antenna $j$, let $g_{i,j}$ denote the $j$-th entry of the effective channel vector $\mathbf{g}_i$. At any sampling instant, each responding tag is in either the absorbing or reflecting load state. We collect these instantaneous load states in $\mathbf{s}=[s_1,\ldots,s_R]$, where $s_i=0$ represents the absorbing state and $s_i=1$ represents the reflecting state of tag $i$. For a given load-state vector $\mathbf{s}$, the corresponding noiseless constellation center at antenna $j$ is $\mu_j(\mathbf{s}) = L_j + \sum_{i=1}^{R} g_{i,j}s_i$.

Since the load state of each tag can take two values, the number of possible constellation centers grows exponentially with the number of responding tags. One tag produces at most two centers, while two tags produce four centers from the four joint load-state combinations. More generally, $R$ responding tags produce up to $2^R$ centers in the I/Q plane, with noise spreading samples around these centers into cluster regions. Since an FM0 response has finite length, not every joint load-state combination may appear, so fewer than $2^R$ regions may be observed. Fig.~\ref{fig:iq_constellation} illustrates this growth from the no tag case to slots with up to four responding tags.

This exponential growth makes explicit constellation extraction increasingly fragile as the collision size grows. The center locations depend on the effective tag channels and carrier leakage. Unfavorable channel geometry can place constellation centers close together, while noise can make the corresponding regions overlap, leaving cluster estimation ambiguous \cite{jin2018parallelbackscatter}.

These limitations motivate a decoding approach that operates on the sampled I/Q waveform itself rather than on estimated cluster centers. Such an approach can use self-attention to jointly model the geometric information created by the simultaneous tag load states and the temporal constraints imposed by FM0 transitions across the RN16 response. This is the basis for the proposed method, which recovers tag responses from the collided waveform without pilot symbols, tag modifications, or a separate cluster extraction stage.

\section{Self-Attention Tag Recovery}
\label{sec:proposed_approach}

This section presents Self-Attention Tag Recovery (SATR), the proposed decoding algorithm for multi-tag collision recovery. SATR operates on the baseband samples collected from a collided slot and recovers the RN16 sequences of the responding tags. We first define the attention operations that serve as common building blocks and then describe the FM0-aware input representation, waveform encoder, and candidate extraction module. We then present the output heads for tag presence, auxiliary channel estimation, and RN16 bit decoding, followed by the permutation-based training objective and the inference procedure used to estimate the number of responding tags and return the decoded RN16 sequences.

\begin{figure*}[t]
\centering
\resizebox{\textwidth}{!}{
\definecolor{tilegray}{RGB}{120,120,120}%
\definecolor{figcream}{HTML}{FFF8E8}%
\begin{tikzpicture}[
  >=Latex, font=\footnotesize, line cap=round,
  tile/.style={draw=figcolor2!70, fill=figcolor2!16, minimum width=0.26cm, minimum height=0.26cm, inner sep=0pt},
  memtile/.style={draw=figcolor2!90, fill=figcolor2!35, minimum width=0.26cm, minimum height=0.26cm, inner sep=0pt},
  cand/.style={draw, minimum width=0.30cm, minimum height=0.30cm, inner sep=0pt, rounded corners=1pt},
  module/.style={draw, semithick, rounded corners=3pt, align=center, inner sep=4pt},
  headbox/.style={draw=figcolor3!85, fill=figcolor3!12, semithick, rounded corners=3pt, align=center, inner sep=3.5pt, minimum width=1.9cm, minimum height=0.62cm},
  eqbox/.style={draw=gray!75, fill=white, rounded corners=2pt, align=center, inner sep=3pt, font=\footnotesize},
  conn/.style={->, line width=0.55pt},
  scoreptr/.style={->, densely dashed, line width=0.45pt},
  stage/.style={font=\footnotesize\bfseries, inner sep=1pt}
]
\node[stage, text=figcolor1!85] at (1.75, 2.55) {collided response};
\node[stage, text=figcolor5!85!black] at (5.95, 2.55) {FM0-aware tokens};
\node[stage, text=figcolor2!95] at (11.1, 2.55) {joint encoding of the response};
\node[stage, text=figcolor4!85] at (15.85, 2.55) {tag candidate formation};
\node[stage, text=figcolor3!85] at (19.55, 2.55) {output heads};

\begin{scope}[shift={(0.15,-0.5)}]
  \draw[gray!60, line width=0.4pt] (0,-0.15) -- (0,1.55);
  \draw[gray!60, line width=0.4pt] (0,-0.15) -- (3.0,-0.15);
  \draw[figcolor1!75, line width=0.5pt, densely dotted]
    (0,0.80)--(0.36,0.80)--(0.36,0)--(0.72,0)--(0.72,0.80)--(1.44,0.80)--(1.44,0)--(1.80,0)--(1.80,0.80)--(2.16,0.80)--(2.16,0)--(2.88,0);
  \draw[figcolor3!80, line width=0.5pt, densely dotted]
    (0,0.48)--(0.72,0.48)--(0.72,0)--(1.08,0)--(1.08,0.48)--(1.44,0.48)--(1.44,0)--(2.16,0)--(2.16,0.48)--(2.52,0.48)--(2.52,0)--(2.88,0);
  \draw[black!80, line width=0.9pt]
    (0,1.28)--(0.36,1.28)--(0.36,0.48)--(0.72,0.48)--(0.72,0.80)--(1.08,0.80)--(1.08,1.28)--(1.44,1.28)--(1.44,0)--(1.80,0)--(1.80,0.80)--(2.16,0.80)--(2.16,0.48)--(2.52,0.48)--(2.52,0)--(2.88,0);
  \foreach \x/\y in {0.18/1.28, 0.54/0.48, 0.90/0.80, 1.26/1.28, 1.62/0, 1.98/0.80, 2.34/0.48, 2.70/0}
    \fill[figcolor2] (\x,\y) circle (1.1pt);
  \node[anchor=north, font=\scriptsize] at (1.44,-0.32) {$\mathbf{y}_n=\mathbf{y}(nT/2)$};
  \draw[figcolor1!75, line width=0.5pt, densely dotted] (0.0,-1.0) -- (0.32,-1.0);
  \node[anchor=west, font=\scriptsize, text=figcolor1!80, inner sep=1pt] at (0.36,-1.0) {tag 1};
  \draw[figcolor3!80, line width=0.5pt, densely dotted] (1.15,-1.0) -- (1.47,-1.0);
  \node[anchor=west, font=\scriptsize, text=figcolor3!80, inner sep=1pt] at (1.51,-1.0) {tag 2};
  \draw[black!80, line width=0.9pt] (2.3,-1.0) -- (2.62,-1.0);
  \node[anchor=west, font=\scriptsize, text=black!80, inner sep=1pt] at (2.66,-1.0) {sum};
\end{scope}

\begin{scope}[shift={(4.55,0)}]
  \foreach \i in {0,...,6}
    \node[tile] (t\i) at (0.30*\i,0) {};
  \node[font=\small] at (2.35,0) {$\cdots$};
  \node[tile] (tlast) at (2.7,0) {};
  \node[anchor=north, font=\scriptsize] at (1.35,-0.22) {tokens $\mathbf{x}_0,\ldots,\mathbf{x}_{31}$};
  \draw[figcolor5, line width=0.7pt, rounded corners=1pt] (0.45,-0.17) rectangle (1.05,0.17);
  \draw[decorate, decoration={brace, amplitude=2.5pt}, figcolor5]
    (0.45,0.22) -- (1.05,0.22) node[midway, above=2pt, font=\scriptsize, text=figcolor5!90!black] {bit $k$};
  \node[draw=figcolor5!85, fill=figcolor5!10, rounded corners=2pt, align=center, font=\scriptsize, inner sep=2.5pt] (zoom) at (1.55,1.2)
    {$\mathbf{y}_{2k},\,\mathbf{y}_{2k+1}\ \rightarrow\ \mathbf{y}^{(+)},\,\mathbf{y}^{(-)},\ n\bmod 2$};
  \draw[figcolor5!70, line width=0.4pt] (0.75,0.30) -- (zoom.south);
\end{scope}
\draw[conn] (3.5,0) -- (4.35,0);

\node[module, draw=figcolor2!85, fill=figcolor2!12, minimum height=0.85cm] (enc) at (9.2,0)
  {waveform encoder\\$L_e$ layers};
\draw[conn] (7.5,0) -- (enc.west);

\begin{scope}[shift={(11.2,0)}]
  \foreach \i in {0,...,6}
    \node[memtile] (m\i) at (0.30*\i,0) {};
  \node[font=\small] at (2.35,0) {$\cdots$};
  \node[memtile] (mlast) at (2.7,0) {};
  \node[anchor=south, font=\scriptsize] at (1.35,0.22) {waveform memory $\mathbf{E}$};
\end{scope}
\draw[conn] (enc.east) -- (11.05,0);

\node[module, draw=figcolor4!85, fill=figcolor4!10, minimum height=0.85cm, minimum width=2.2cm] (cem) at (15.85,0)
  {candidate extraction\\$L_c$ layers};
\draw[conn] (14.15,0) -- (cem.west);

\begin{scope}[shift={(14.2,-1.45)}]
  \node[cand, draw=figcolor1!80, fill=figcolor1!25] at (-0.54,0) {};
  \node[cand, draw=figcolor3!80, fill=figcolor3!25] at (-0.18,0) {};
  \node[cand, draw=figcolor5!90, fill=figcolor5!30] at (0.18,0) {};
  \node[cand, draw=tilegray, fill=tilegray!25] at (0.54,0) {};
  \node[anchor=north, font=\scriptsize, align=center] at (0,-0.28) {$M$ trainable candidates $\mathbf{Z}^{(0)}$};
\end{scope}
\draw[conn] (14.9,-1.2) -- (15.55,-0.55);

\begin{scope}[shift={(17.75,0)}]
  \node[cand, draw=figcolor1!80, fill=figcolor1!45] (z1) at (0,0.55) {};
  \node[cand, draw=figcolor3!80, fill=figcolor3!45] (z2) at (0,0.15) {};
  \node[cand, draw=figcolor5!90, fill=figcolor5!55] (z3) at (0,-0.25) {};
  \node[cand, draw=tilegray, fill=tilegray!40] (z4) at (0,-0.65) {};
  \node[anchor=south, font=\scriptsize, align=center] at (-0.45,0.82) {candidate\\representations $\mathbf{z}_i$};
\end{scope}
\draw[conn] (cem.east) -- (17.55,0);

\node[headbox] (ph) at (19.55,1.6) {presence head};
\node[headbox] (ch) at (19.55,-0.35) {channel head};
\node[headbox] (bd) at (19.55,-2.3) {bit decoder\\[-1pt]{\scriptsize $L_b$ layers}};
\draw[conn, figcolor2!90, rounded corners=3pt] (12.1,-0.15) -- (12.1,-3.05) -- (19.55,-3.05) -- (bd.south);
\node[font=\footnotesize, text=figcolor2!90, fill=white, inner sep=1.5pt] at (16.6,-3.05) {$\mathbf{E}$};
\draw[conn] (18.0,0.55) .. controls (18.35,0.75) and (18.45,1.3) .. (ph.west);
\draw[conn] (18.0,0.15) .. controls (18.35,-0.1) .. (ch.west);
\draw[conn] (18.0,-0.45) .. controls (18.35,-1.9) .. (bd.west);

\begin{scope}[shift={(21.15,1.25)}]
  \draw[figcolor1!85, fill=figcolor1!40] (0,0.72) rectangle (1.455,0.86);
  \draw[figcolor3!85, fill=figcolor3!40] (0,0.48) rectangle (1.395,0.62);
  \draw[figcolor5!95, fill=figcolor5!30] (0,0.24) rectangle (0.105,0.38);
  \draw[tilegray,      fill=tilegray!30]  (0,0.00) rectangle (0.06,0.14);
  \draw[black!70, densely dashed, line width=0.5pt] (0.75,-0.10) -- (0.75,0.98) node[above, font=\scriptsize] {$p_{\mathrm{th}}$};
  \node[font=\scriptsize, anchor=west] at (1.55,0.79) {$\hat{p}_1=0.97$};
  \node[font=\scriptsize, anchor=west] at (1.55,0.55) {$\hat{p}_2=0.93$};
  \node[font=\scriptsize, anchor=west, text=tilegray] at (1.55,0.31) {$\hat{p}_3=0.07$};
  \node[font=\scriptsize, anchor=west, text=tilegray] at (1.55,0.07) {$\hat{p}_4=0.04$};
  \node[font=\scriptsize\bfseries, anchor=west, text=figcolor3!50!black] at (0.0,-0.38) {$\hat{R}=2$ tags detected};
\end{scope}
\begin{scope}[shift={(21.15,-0.9)}]
  \draw[gray!60, line width=0.4pt] (0,0) -- (1.1,0) node[right=0pt, font=\scriptsize, text=gray] {I};
  \draw[gray!60, line width=0.4pt] (0,0) -- (0,1.1) node[above=0pt, font=\scriptsize, text=gray] {Q};
  \fill[figcolor1] (0.35,0.75) circle (1.4pt);
  \fill[figcolor3] (0.85,0.30) circle (1.4pt);
  \node[font=\scriptsize, anchor=west] at (1.35,0.55) {$\hat{\mathbf{g}}_i$};
\end{scope}
\begin{scope}[shift={(21.15,-1.95)}]
  \node[font=\scriptsize, anchor=west, text=figcolor1!85] at (0,0.30) {\texttt{0101101}$\cdots$\texttt{0}};
  \node[font=\scriptsize, anchor=west, text=figcolor3!85] at (0,0.02) {\texttt{1010011}$\cdots$\texttt{1}};
  \node[font=\scriptsize, anchor=west, text=tilegray] at (0,-0.26) {not present};
  \node[font=\scriptsize, anchor=west, text=tilegray] at (0,-0.54) {not present};
  \node[font=\scriptsize, anchor=west] at (0,-0.92) {decoded RN16 $\hat{\mathbf{b}}_i$};
\end{scope}

\begin{scope}[on background layer]
  \path[fill=figcream, draw=gray!35, line width=0.5pt, rounded corners=5pt]
    ([shift={(-8pt,-8pt)}]current bounding box.south west)
    rectangle
    ([shift={(8pt,8pt)}]current bounding box.north east);
\end{scope}

\end{tikzpicture}}
\caption{Overview of the proposed SATR algorithm, shown for a collision of two tags. The waveform memory $\mathbf{E}$ is read by the candidate extraction module and by the bit decoder. Two of the $M=4$ candidates exceed the presence threshold, and the decoded RN16 sequences of the two detected tags are returned.}
\label{fig:star_arch}
\vspace{-0.5cm}
\end{figure*}
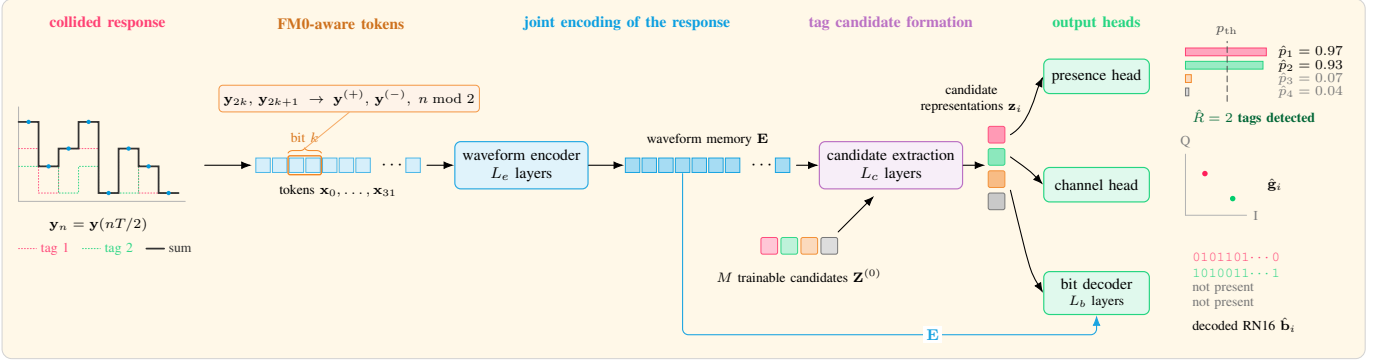

Fig.~\ref{fig:star_arch} summarizes the main processing flow of SATR. The sampled RN16 waveform is first converted into FM0-aware tokens, one for each half-symbol sample. These tokens combine the raw complex samples with features derived from the two half-symbol samples of each RN16 bit and with positional information. A self-attention waveform encoder transforms the token sequence into the waveform memory $\mathbf{E}$, a contextual representation of the received slot used for candidate extraction and bit decoding. The candidate extraction module begins with $M$ trainable vectors representing possible responding tags and refines them by attending to $\mathbf{E}$, yielding waveform-dependent candidate representations $\mathbf{z}_i$. The output heads, i.e., task-specific output modules, use each $\mathbf{z}_i$ as follows: the presence head estimates whether the candidate is present, the channel head predicts an auxiliary effective channel for candidate assignment during training, and the bit decoder reconstructs the corresponding RN16 sequence.

\subsection{Preliminaries}
\label{subsec:satr_attention}
Before describing the individual processing stages in detail, we define the attention and related operations used throughout SATR, following the standard transformer architecture \cite{vaswani2017attention}.

\paragraph{Attention}
Attention allows each query to gather relevant information from a set of available representations. A query represents the position being updated, each key is compared with that query to determine relevance, and its corresponding value provides the information used in the update. Let $\mathbf Q$, $\mathbf K$, and $\mathbf V$ collect these queries, keys, and values as rows, respectively. Scaled dot product attention is computed as
\begin{equation}
	\mathbf{S}
	=
	\operatorname{softmax}\!\left(
	\frac{\mathbf{Q}\mathbf{K}^{\mathsf{T}}}{\sqrt{d_h}}
	\right),
	\qquad
	\operatorname{Attn}(\mathbf{Q},\mathbf{K},\mathbf{V})
	=
	\mathbf{S}\mathbf{V},
	\label{eq:attention}
\end{equation}
where $d_h$ is the common dimension of the query, key, and value representations, and the softmax is applied across each row. Each row of $\mathbf{S}$ gives the attention weights assigned by one query to the value representations, so the corresponding output row is their weighted combination.

\paragraph{Multihead attention}
Multihead attention applies the attention operation through $n_h$ parallel heads, which can capture complementary relationships. Let $\mathbf{A}$ denote the query sequence and $\mathbf{C}$ the context sequence, both consisting of representations of dimension $d$. Each head $r\in\{1,\ldots,n_h\}$ uses separate projections to obtain
\begin{equation}
	\mathbf{Q}^{(r)} = \mathbf{A}\mathbf{W}^{(r)}_{Q}, \quad
	\mathbf{K}^{(r)} = \mathbf{C}\mathbf{W}^{(r)}_{K}, \quad
	\mathbf{V}^{(r)} = \mathbf{C}\mathbf{W}^{(r)}_{V},
	\label{eq:common_attention_proj}
\end{equation}
where $\mathbf{W}^{(r)}_{Q},\mathbf{W}^{(r)}_{K},\mathbf{W}^{(r)}_{V}\in\R^{d\times d_h}$ are trainable projection matrices and $d_h=d/n_h$. The output of head $r$ is obtained using the attention operation in \eqref{eq:attention} as
\begin{equation}
	\mathbf{H}^{(r)}
	=
	\operatorname{Attn}
	\bigl(\mathbf{Q}^{(r)},\mathbf{K}^{(r)},\mathbf{V}^{(r)}\bigr).
	\label{eq:head_output}
\end{equation}
Finally, the head outputs are concatenated along the feature dimension to form $\mathbf{H}=[\mathbf{H}^{(1)},\ldots,\mathbf{H}^{(n_h)}]$. An output projection then gives
\begin{equation}
	\operatorname{MHA}_{\Theta}(\mathbf{A},\mathbf{C})
	=
	\mathbf{H}\mathbf{W}_{O},
	\label{eq:common_mha}
\end{equation}
where $\mathbf{W}_{O}\in\R^{d\times d}$ is a trainable projection matrix and $\Theta$ collects all trainable parameters of the operation. Bias terms are omitted from the notation for clarity.

\paragraph{Self-attention}
Self-attention uses the same sequence as both the query and the context. Each position can therefore incorporate information from every position in the sequence, allowing relationships across the sequence to be represented. For a sequence $\mathbf{A}$, self-attention is defined as
\begin{equation}
	\operatorname{SA}_{\Theta}(\mathbf{A})
	=
	\operatorname{MHA}_{\Theta}(\mathbf{A},\mathbf{A}).
	\label{eq:common_self_attention}
\end{equation}

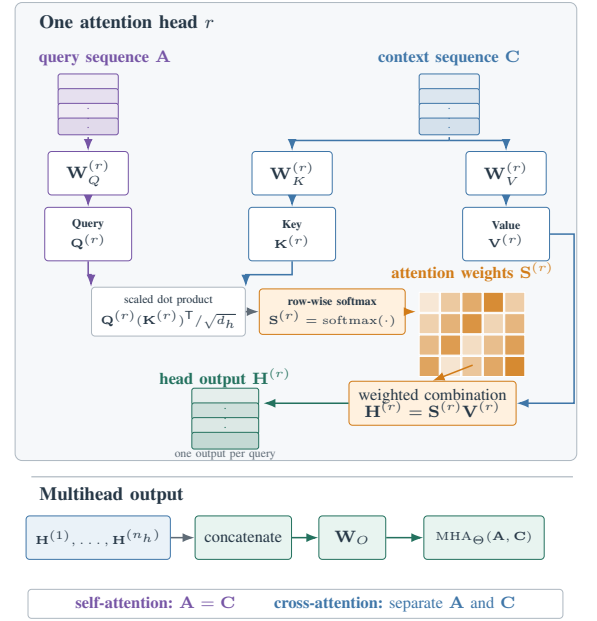
\begin{figure}[t]
	\centering
	\resizebox{\columnwidth}{!}{
\definecolor{ink}{HTML}{263746}
\definecolor{muted}{HTML}{5F6B76}
\definecolor{linegray}{HTML}{AEB8C1}
\definecolor{panel}{HTML}{F6F8FA}
\definecolor{query}{HTML}{7655A5}
\definecolor{querylight}{HTML}{F0EBF7}
\definecolor{context}{HTML}{3F76A8}
\definecolor{contextlight}{HTML}{EAF2F8}
\definecolor{weight}{HTML}{D17B17}
\definecolor{weightlight}{HTML}{FFF1DF}
\definecolor{output}{HTML}{23715E}
\definecolor{outputlight}{HTML}{E8F4F0}
\begin{tikzpicture}[
	>=Latex,
	flow/.style={-{Latex[length=1.8mm]},draw=muted,line width=0.65pt},
	thinflow/.style={-{Latex[length=1.5mm]},draw=linegray,line width=0.55pt},
	op/.style={
		draw=linegray,
		fill=white,
		rounded corners=1.5pt,
		minimum height=6.5mm,
		align=center,
		font=\scriptsize,
		text=ink,
		inner xsep=4pt
	},
	qop/.style={op,draw=query},
	cop/.style={op,draw=context},
	wop/.style={op,draw=weight,fill=weightlight},
	outop/.style={op,draw=output,fill=outputlight},
]

\node[draw=linegray,fill=panel,rounded corners=3pt,
	minimum width=84mm,minimum height=68.5mm,anchor=north west]
	(panelbox) at (0,9.55) {};
\node[font=\footnotesize\bfseries,text=ink,anchor=west]
	at (0.25,9.25) {One attention head $r$};

\node[font=\scriptsize\bfseries,text=query] at (1.35,8.72)
	{query sequence $\mathbf A$};
\node[font=\scriptsize\bfseries,text=context] at (6.50,8.72)
	{context sequence $\mathbf C$};

\foreach \y/\shade in {8.35/10,8.13/18,7.91/12,7.69/22}
	\node[draw=query,fill=query!\shade!white,rounded corners=1pt,
		minimum width=9mm,minimum height=2.5mm,inner sep=0pt]
		at (1.10,\y) {};
\node[font=\tiny,text=query] at (1.10,7.92) {$\vdots$};

\foreach \y/\shade in {8.35/10,8.13/18,7.91/12,7.69/22}
	\node[draw=context,fill=context!\shade!white,rounded corners=1pt,
		minimum width=9mm,minimum height=2.5mm,inner sep=0pt]
		at (6.50,\y) {};
\node[font=\tiny,text=context] at (6.50,7.92) {$\vdots$};

\node[qop,minimum width=12mm] (wq) at (1.10,6.99)
	{$\mathbf W_Q^{(r)}$};
\node[cop,minimum width=12mm] (wk) at (4.15,6.99)
	{$\mathbf W_K^{(r)}$};
\node[cop,minimum width=12mm] (wv) at (7.35,6.99)
	{$\mathbf W_V^{(r)}$};
\draw[flow,draw=query] (1.10,7.53) -- (wq.north);
\coordinate (cbranch) at (6.50,7.48);
\draw[draw=context,line width=0.65pt] (6.50,7.53) -- (cbranch);
\draw[flow,draw=context] (cbranch) -| (wk.north);
\draw[flow,draw=context] (cbranch) -| (wv.north);

\node[qop,minimum width=13mm,minimum height=8mm,font=\tiny]
	(qmat) at (1.10,6.08)
	{\textbf{Query}\\[2pt] $\mathbf Q^{(r)}$};
\node[cop,minimum width=13mm,minimum height=8mm,font=\tiny]
	(kmat) at (4.15,6.08)
	{\textbf{Key}\\[2pt] $\mathbf K^{(r)}$};
\node[cop,minimum width=13mm,minimum height=8mm,font=\tiny]
	(vmat) at (7.35,6.08)
	{\textbf{Value}\\[2pt] $\mathbf V^{(r)}$};
\draw[flow,draw=query] (wq) -- (qmat);
\draw[flow,draw=context] (wk) -- (kmat);
\draw[flow,draw=context] (wv) -- (vmat);

\node[op,minimum width=22mm,font=\tiny] (score) at (2.30,4.92)
	{scaled dot product\\[2pt]
	$\mathbf Q^{(r)}(\mathbf K^{(r)})^{\mathsf T}/\sqrt{d_h}$};
\node[wop,minimum width=18mm,font=\tiny] (softmax) at (4.75,4.92)
	{\textbf{row-wise softmax}\\[2pt]
	$\mathbf S^{(r)}=\operatorname{softmax}(\cdot)$};
\draw[flow,draw=query] (qmat.south) -- ++(0,-0.20) -| (score.north west);
\draw[flow,draw=context] (kmat.south) -- ++(0,-0.20) -| (score.north east);
\draw[flow] (score) -- (softmax);

\node[font=\scriptsize\bfseries,text=weight] at (6.85,5.51)
	{attention weights $\mathbf S^{(r)}$};
\foreach \row in {0,...,3}{
	\foreach \col in {0,...,4}{
		\pgfmathtruncatemacro{\shade}{18+mod(17*\row+23*\col,72)}
		\node[minimum width=3.0mm,minimum height=3.0mm,inner sep=0pt,
			draw=white,line width=0.25pt,fill=weight!\shade!white]
			at ({6.20+0.32*\col},{5.06-0.32*\row}) {};
	}
}
\draw[flow,draw=weight] (softmax) -- (6.00,4.92);

\node[wop,minimum width=24mm] (combine) at (6.25,3.55)
	{weighted combination\\[-1pt]
	$\mathbf H^{(r)}=\mathbf S^{(r)}\mathbf V^{(r)}$};
\draw[flow,draw=weight] (6.85,4.12) -- (combine.north);
\draw[flow,draw=context] (vmat.east) -- ++(0.36,0) |- (combine.east);

\node[font=\scriptsize\bfseries,text=output] at (3.15,3.93)
	{head output $\mathbf H^{(r)}$};
\foreach \y/\shade in {3.65/12,3.43/22,3.21/16,2.99/27}
	\node[draw=output,fill=output!\shade!white,rounded corners=1pt,
		minimum width=10mm,minimum height=2.5mm,inner sep=0pt]
		at (3.15,\y) {};
\node[font=\tiny,text=output] at (3.15,3.43) {$\vdots$};
\draw[flow,draw=output] (combine) -- (3.72,3.55);
\node[font=\tiny,text=muted] at (3.15,2.78)
	{one output per query};

\draw[linegray,line width=0.5pt] (0.25,2.45) -- (8.35,2.45);
\node[font=\footnotesize\bfseries,text=ink,anchor=west]
	at (0.25,2.17) {Multihead output};

\node[op,fill=contextlight,draw=context,minimum width=18mm,font=\tiny]
	(allheads) at (1.25,1.54)
	{$\mathbf H^{(1)},\ldots,\mathbf H^{(n_h)}$};
\node[outop,minimum width=13mm] (concat) at (3.42,1.54)
	{concatenate};
\node[outop,minimum width=10mm] (wo) at (5.05,1.54)
	{$\mathbf W_O$};
\node[outop,minimum width=18mm,font=\tiny] (mha) at (7.02,1.54)
	{$\operatorname{MHA}_{\Theta}
	(\mathbf A,\mathbf C)$};
\draw[flow] (allheads) -- (concat);
\draw[flow,draw=output] (concat) -- (wo);
\draw[flow,draw=output] (wo) -- (mha);

\node[draw=linegray,fill=white,rounded corners=1.5pt,
	font=\scriptsize,text=ink,minimum width=80mm,
	inner xsep=4pt,inner ysep=3pt] at (4.20,0.55)
	{\textcolor{query}{\textbf{self-attention:} $\mathbf A=\mathbf C$}
	\qquad
	\textcolor{context}{\textbf{cross-attention:} separate $\mathbf A$ and $\mathbf C$}};

\end{tikzpicture}}
	\caption{General attention operation for head $r$ and the resulting multihead output. Self-attention uses $\mathbf{A}=\mathbf{C}$, whereas cross-attention uses separate query and context sequences.}
	\label{fig:general_attention}
	\vspace{-0.75cm}
\end{figure}

\paragraph{Cross-attention}
Cross-attention updates a query sequence using information from a separate context sequence. The query sequence determines the positions to be updated, while the context sequence supplies the information incorporated into those positions. For query sequence $\mathbf{A}$ and context sequence $\mathbf{C}$, cross-attention is defined as
\begin{equation}
	\operatorname{CA}_{\Theta}(\mathbf{A},\mathbf{C})
	=
	\operatorname{MHA}_{\Theta}(\mathbf{A},\mathbf{C}).
	\label{eq:common_cross_attention}
\end{equation}
Fig.~\ref{fig:general_attention} summarizes the attention mechanism used throughout SATR.

\paragraph{Feedforward network and layer normalization}
After attention exchanges information across sequence positions, a feedforward network refines each representation independently. For a sequence $\mathbf{A}$, the transformation is given by
\begin{equation}
	\operatorname{FFN}_{\Theta}(\mathbf{A})
	=
	\operatorname{ReLU}\!\left(
	\mathbf{A}\mathbf{W}^{\mathrm{ff}}_{1}
	+
	\mathbf{b}^{\mathrm{ff}}_{1}
	\right)
	\mathbf{W}^{\mathrm{ff}}_{2}
	+
	\mathbf{b}^{\mathrm{ff}}_{2},
	\label{eq:common_ffn}
\end{equation}
where $d_{\mathrm{ff}}$ is the hidden dimension and $\operatorname{ReLU}(\cdot)$ is applied elementwise. The trainable parameters are $\mathbf{W}^{\mathrm{ff}}_{1}\in\R^{d\times d_{\mathrm{ff}}}$, $\mathbf{W}^{\mathrm{ff}}_{2}\in\R^{d_{\mathrm{ff}}\times d}$, $\mathbf{b}^{\mathrm{ff}}_{1}\in\R^{d_{\mathrm{ff}}}$, and $\mathbf{b}^{\mathrm{ff}}_{2}\in\R^{d}$. Layer normalization is denoted by $\operatorname{LN}(\cdot)$. It normalizes the features of each representation independently and applies trainable scale and offset parameters.

\subsection{FM0-Aware Input Representation}
\label{subsec:satr_input}
The FM0-aware input representation converts the collided RN16 waveform into real-valued token features while preserving its I/Q geometry and temporal structure. Each complex sample contains the combined load-state contributions of the responding tags. SATR samples the waveform once per FM0 half symbol. With sampling period $T_{\mathrm{s}} = T/2$, where $T$ is the FM0 symbol period, the resulting samples are
\begin{equation}
	\mathbf{y}_n = \mathbf{y}(nT_{\mathrm{s}}) \in \C^{N_{\mathrm{rx}}}, \qquad n = 0,\dots,31.
	\label{eq:star_sampling}
\end{equation}
The samples $\mathbf{y}_{2k}$ and $\mathbf{y}_{2k+1}$ form the pair associated with RN16 bit $k$. In FM0, the presence or absence of a midpoint transition determines the bit value, while a transition occurs at every symbol boundary. SATR therefore retains the raw samples in temporal order to preserve the transitions between bit intervals and forms their sum and difference within each pair to make the midpoint behavior explicit. For each RN16 bit position $k$, the sum and difference features are defined as
\begin{equation}
	\begin{aligned}
	\mathbf{y}^{(+)}_n &=
	\begin{cases}
		\mathbf{y}_{2k+1} + \mathbf{y}_{2k}, & n = 2k+1,\\
		\mathbf{0}, & n = 2k,
	\end{cases}\\
	\mathbf{y}^{(-)}_n &=
	\begin{cases}
		\mathbf{y}_{2k+1} - \mathbf{y}_{2k}, & n = 2k+1,\\
		\mathbf{0}, & n = 2k.
	\end{cases}
	\end{aligned}
	\label{eq:star_sumdiff}
\end{equation}
Note that the features $\mathbf{y}^{(+)}_n$ and $\mathbf{y}^{(-)}_n$ are formed only from the two half-symbol samples of the same bit interval and not across adjacent bit intervals. The entries at $n=2k$ are set to zero so that all sample tokens have the same dimension.

Using the raw samples and the half-symbol pair features, SATR constructs a real-valued feature vector for each sample position. The real and imaginary parts are concatenated so that the complex waveform can be processed by standard real-valued neural network layers. The resulting vector is
\begin{align}
	\mathbf{u}_n =
	\big[&
		(\operatorname{Re}\{\mathbf{y}_n\})^{\mathsf{T}},
		(\operatorname{Im}\{\mathbf{y}_n\})^{\mathsf{T}},
		(\operatorname{Re}\{\mathbf{y}^{(+)}_n\})^{\mathsf{T}},
		(\operatorname{Im}\{\mathbf{y}^{(+)}_n\})^{\mathsf{T}}, \nonumber\\
	&	(\operatorname{Re}\{\mathbf{y}^{(-)}_n\})^{\mathsf{T}},
		(\operatorname{Im}\{\mathbf{y}^{(-)}_n\})^{\mathsf{T}},
		n \bmod 2
	\big]^{\mathsf{T}},
	\label{eq:star_input}
\end{align}
where $\mathbf{u}_n\in\R^{6N_{\mathrm{rx}}+1}$ and the final entry $n \bmod 2$ indicates whether sample $n$ corresponds to the first or second half symbol of its RN16 bit interval. This position information helps the encoder interpret the raw sample and the pair features.
Each feature vector $\mathbf{u}_n$ is mapped to a $d$-dimensional token embedding by a shared trainable linear layer. A fixed sinusoidal positional encoding $\mathbf{p}_n\in\R^{d}$ is then added to encode the sample location within the RN16 response, i.e.,
\begin{equation}
	\mathbf{x}_n = \mathbf{W}_{\mathrm{in}}\mathbf{u}_n + \mathbf{b}_{\mathrm{in}} + \mathbf{p}_n \in \R^{d} ,
	\label{eq:star_embed}
\end{equation}
where $\mathbf{W}_{\mathrm{in}}\in\R^{d\times(6N_{\mathrm{rx}}+1)}$ and $\mathbf{b}_{\mathrm{in}}\in\R^{d}$ are trainable parameters. The token sequence $\mathbf{X} = [\mathbf{x}_0,\dots,\mathbf{x}_{31}]^{\mathsf{T}}\in\R^{32\times d}$ forms the input to the waveform encoder.

\subsection{Waveform Encoder}
\label{subsec:satr_encoder}
The waveform encoder transforms the embedded sequence $\mathbf{X}$ into the waveform memory $\mathbf{E}$, retaining one representation for each sample position. Self-attention allows each position to incorporate information from the full received waveform. This global context is useful because the effective channels remain fixed over the RN16 response, so repeated joint load states produce samples around the same constellation centers across the slot, while transitions between the two half symbols of each bit determine the FM0 bit value.

The encoder contains $L_e$ layers, each applying self-attention followed by a feedforward network. Layer normalization is applied before each operation, and a residual connection adds each output to its input. The input to the first layer is $\mathbf{E}^{(0)}=\mathbf{X}$, and $\mathbf{E}^{(\ell)}\in\R^{32\times d}$ denotes the output of layer $\ell$. For each layer, the self-attention update is given as
\begin{equation}
	\tilde{\mathbf{E}}^{(\ell)}
	=
	\mathbf{E}^{(\ell-1)}
	+
	\operatorname{SA}_{\Theta_{\mathrm{enc,sa}}^{(\ell)}}
	\!\left(
	\operatorname{LN}\!\left(\mathbf{E}^{(\ell-1)}\right)
	\right),
	\label{eq:enc_mha}
\end{equation}
where $\tilde{\mathbf{E}}^{(\ell)}\in\R^{32\times d}$ is the sequence after the residual self-attention step. The feedforward update is then written as
\begin{equation}
	\mathbf{E}^{(\ell)}
	=
	\tilde{\mathbf{E}}^{(\ell)}
	+
	\operatorname{FFN}_{\Theta_{\mathrm{enc,ff}}^{(\ell)}}
	\!\left(
	\operatorname{LN}\!\left(\tilde{\mathbf{E}}^{(\ell)}\right)
	\right),
	\label{eq:enc_ffn}
\end{equation}
where $\Theta_{\mathrm{enc,sa}}^{(\ell)}$ and $\Theta_{\mathrm{enc,ff}}^{(\ell)}$ denote the trainable parameters of the self-attention and feedforward operations in encoder layer $\ell$, respectively. After the final encoder layer, the waveform memory is written as
\begin{equation}
	\mathbf{E}
	=
	\mathbf{E}^{(L_e)}
	=
	[\mathbf{e}_0,\ldots,\mathbf{e}_{31}]^{\mathsf{T}}
	\in\R^{32\times d}.
	\label{eq:waveform_memory}
\end{equation}
Each vector $\mathbf{e}_n$ remains aligned with sample position $n$ while incorporating information from the full received slot.

\subsection{Candidate Extraction}
\label{subsec:satr_candidates}
The candidate extraction module maps the waveform memory to $M$ candidate tag representations. The fixed number of candidates allows SATR to perform set prediction \cite{carion2020detr} without knowing the number of responding tags in advance, and the outputs are not tied to any physical tag order. The module begins with trainable candidate vectors that are shared across received slots, given by
\begin{equation}
	\mathbf{Z}^{(0)}
	=
	[\mathbf{z}^{(0)}_1,\dots,\mathbf{z}^{(0)}_M]^{\mathsf{T}}
	\in\R^{M\times d},
	\label{eq:init_candidates}
\end{equation}
where the $i$-th row is the initial representation of candidate $i$.

The candidate extraction module contains $L_c$ layers, each using the same normalization and residual structure as the waveform encoder. The output of extraction layer $\ell$ is denoted by $\mathbf{Z}^{(\ell)}\in\R^{M\times d}$, with $\mathbf{Z}^{(0)}$ used as the input to the first layer. Each layer first applies self-attention across the candidate representations, allowing the candidates to exchange information before they attend to the waveform memory. This interaction allows the candidates to coordinate their predictions rather than independently representing the same response. For $\ell=1,\ldots,L_c$, the residual self-attention update is given by
\begin{equation}
	\tilde{\mathbf{Z}}^{(\ell)}
	=
	\mathbf{Z}^{(\ell-1)}
	+
	\operatorname{SA}_{\Theta_{\mathrm{ext,sa}}^{(\ell)}}
	\!\left(
	\operatorname{LN}\!\left(\mathbf{Z}^{(\ell-1)}\right)
	\right).
	\label{eq:dec_self}
\end{equation}
Cross-attention then allows each candidate representation to use the waveform memory as context. The candidate representations provide the queries, and the waveform memory provides the keys and values. The residual cross-attention update is written as
\begin{equation}
	\bar{\mathbf{Z}}^{(\ell)}
	=
	\tilde{\mathbf{Z}}^{(\ell)}
	+
	\operatorname{CA}_{\Theta_{\mathrm{ext,ca}}^{(\ell)}}
	\!\left(
	\operatorname{LN}\!\left(\tilde{\mathbf{Z}}^{(\ell)}\right),
	\mathbf{E}
	\right),
	\label{eq:dec_cross}
\end{equation}
where $\bar{\mathbf{Z}}^{(\ell)}\in\R^{M\times d}$ denotes the intermediate candidate sequence after cross-attention. The layer then refines each candidate representation with a feedforward update, given by
\begin{equation}
	\mathbf{Z}^{(\ell)}
	=
	\bar{\mathbf{Z}}^{(\ell)}
	+
	\operatorname{FFN}_{\Theta_{\mathrm{ext,ff}}^{(\ell)}}
	\!\left(
	\operatorname{LN}\!\left(\bar{\mathbf{Z}}^{(\ell)}\right)
	\right),
	\label{eq:dec_ffn}
\end{equation}
where $\Theta_{\mathrm{ext,sa}}^{(\ell)}$, $\Theta_{\mathrm{ext,ca}}^{(\ell)}$, and $\Theta_{\mathrm{ext,ff}}^{(\ell)}$ denote the trainable parameters of the self-attention, cross-attention, and feedforward operations in layer $\ell$, respectively.

The output of the candidate extraction module is
\begin{equation}
	\mathbf{Z}
	=
	\mathbf{Z}^{(L_c)}
	=
	[\mathbf{z}_1,\ldots,\mathbf{z}_{M}]^{\mathsf{T}}
	\in\R^{M\times d},
	\label{eq:decoder_output}
\end{equation}
where $\mathbf{z}_i$ is the final waveform dependent representation of candidate $i$ and summarizes the information associated with one possible responding tag.

\subsection{Output Heads}
\label{subsec:satr_heads}
Three output heads convert each candidate representation $\mathbf{z}_i$ into the final predictions. The presence head estimates whether the candidate is present, the channel head predicts an auxiliary effective channel used only during training, and the bit decoder estimates the RN16 sequence.

The presence head estimates whether candidate $i$ corresponds to a responding tag. It applies a two-layer feedforward network to $\mathbf{z}_i$ and passes the resulting logit $\psi_i$ through a sigmoid to obtain the presence probability $\hat{p}_i$, i.e.,
\begin{equation}
	\psi_i = \operatorname{FFN}_{\Theta_p}(\mathbf{z}_i),
	\qquad
	\hat{p}_i = \sigma(\psi_i),
	\label{eq:presence_head}
\end{equation}
where $\sigma(\cdot)$ is the sigmoid function and $\Theta_p$ denotes the trainable parameters of a two layer feedforward network with hidden dimension $d_{\mathrm{ff}}$ and a scalar output.

The channel head produces the auxiliary effective-channel estimate through the same feedforward form with $2N_{\mathrm{rx}}$ outputs, given by
\begin{equation}
	\big[(\operatorname{Re}\{\hat{\mathbf{g}}_i\})^{\mathsf{T}},(\operatorname{Im}\{\hat{\mathbf{g}}_i\})^{\mathsf{T}}\big]^{\mathsf{T}}
	=
	\operatorname{FFN}_{\Theta_g}(\mathbf{z}_i)
	\in \R^{2N_{\mathrm{rx}}},
	\label{eq:chan_head}
\end{equation}
where $\Theta_g$ denotes the trainable parameters of a two layer feedforward network with hidden dimension $d_{\mathrm{ff}}$ and output dimension $2N_{\mathrm{rx}}$. The first $N_{\mathrm{rx}}$ outputs give the real part of $\hat{\mathbf{g}}_i\in\C^{N_{\mathrm{rx}}}$ and the remaining $N_{\mathrm{rx}}$ the imaginary part. This estimate is used only for training supervision and the assignment cost, not during inference.

The RN16 bit decoder estimates an RN16 sequence for each candidate one bit at a time. This sequential decoding follows FM0 encoding, where the load state entering each bit interval depends on the preceding bits. To predict bit $k$ for candidate $i$, the decoder combines its representation $\mathbf{z}_i$ with a learned position embedding $\mathbf{p}^{\mathrm{b}}_k$ for the current bit position and a previous bit embedding $\mathbf{r}_{i,k}$ for the decoding history. The two embedding tables are shared across candidates. For $k=0$, $\mathbf{r}_{i,0}$ is a learned start embedding. For $k>0$, it represents the true previous bit during training and the previously decoded bit during inference. The decoder input is
\begin{equation}
	\mathbf{t}_{i,k}
	=
	\mathbf{z}_i
	+
	\mathbf{p}^{\mathrm{b}}_k
	+
	\mathbf{r}_{i,k}
	\in\R^d.
	\label{eq:bit_input}
\end{equation}
The input tokens for candidate $i$ are stacked as $\mathbf{T}^{(0)}_i = [\mathbf{t}_{i,0},\ldots,\mathbf{t}_{i,15}]^{\mathsf{T}}\in\R^{16\times d}$. The bit decoder processes this sequence through $L_b$ layers, and $\mathbf{T}^{(\ell)}_i\in\R^{16\times d}$ denotes the output of layer $\ell$. Each operation uses the same normalization and residual structure as the waveform encoder. Within each layer, causal self-attention incorporates information from the current and preceding bit positions without using later positions. Cross-attention then uses the bit representations as queries and the waveform memory $\mathbf{E}$ as context before a feedforward network refines each bit representation. For each layer, the updates are given by
\begin{equation}
	\begin{aligned}
	\tilde{\mathbf{T}}^{(\ell)}_i
	&=
	\mathbf{T}^{(\ell-1)}_i
	+
	\operatorname{SA}^{\mathrm{causal}}_{\Theta_{\mathrm{b,sa}}^{(\ell)}}
	\!\left(\operatorname{LN}(\mathbf{T}^{(\ell-1)}_i)\right),\\
	\bar{\mathbf{T}}^{(\ell)}_i
	&=
	\tilde{\mathbf{T}}^{(\ell)}_i
	+
	\operatorname{CA}_{\Theta_{\mathrm{b,ca}}^{(\ell)}}
	\!\left(\operatorname{LN}(\tilde{\mathbf{T}}^{(\ell)}_i),\mathbf{E}\right),\\
	\mathbf{T}^{(\ell)}_i
	&=
	\bar{\mathbf{T}}^{(\ell)}_i
	+
	\operatorname{FFN}_{\Theta_{\mathrm{b,ff}}^{(\ell)}}
	\!\left(\operatorname{LN}(\bar{\mathbf{T}}^{(\ell)}_i)\right),
	\end{aligned}
	\label{eq:bit_decoder_layer}
\end{equation}
where $\operatorname{SA}^{\mathrm{causal}}$ denotes self-attention restricted to the current and preceding bit positions, and $\Theta_{\mathrm{b,sa}}^{(\ell)}$, $\Theta_{\mathrm{b,ca}}^{(\ell)}$, and $\Theta_{\mathrm{b,ff}}^{(\ell)}$ are the trainable parameters of the self-attention, cross-attention, and feedforward operations in layer $\ell$. After the final decoder layer, the output sequence is $\mathbf{T}^{(L_b)}_i = [\mathbf{h}_{i,0},\ldots,\mathbf{h}_{i,15}]^{\mathsf{T}}\in\R^{16\times d}$, where $\mathbf{h}_{i,k}\in\R^d$ is the hidden representation for bit position $k$ of candidate $i$. A linear map shared across all candidates and bit positions, followed by a sigmoid activation, produces the predicted probability that the $k$-th bit is one, i.e.,
\begin{equation}
	\hat q_i[k]
	=
	\sigma\!\left(
	\mathbf{w}_{\mathrm{b}}^{\mathsf{T}}\mathbf{h}_{i,k}
	+
	b_{\mathrm{b}}
	\right)
	\in(0,1),
\end{equation}
where $\mathbf{w}_{\mathrm{b}}\in\R^d$ and $b_{\mathrm{b}}\in\R$ are trainable parameters. The complete SATR architecture is summarized in Table~\ref{tab:star_arch}.

\begin{table}[t]
	\centering
	\caption{SATR architecture parameters} \vspace{-0.1cm}
	\label{tab:star_arch}
	\scriptsize
	\setlength{\tabcolsep}{2pt}
	\renewcommand{\arraystretch}{1.25}
	\rowcolors{2}{white}{tblzebra}
	\begin{tabularx}{0.95\columnwidth}{@{}>{\raggedright\arraybackslash}p{0.25\columnwidth}>{\raggedright\arraybackslash}X>{\centering\arraybackslash}p{0.22\columnwidth}@{}}
		\rowcolor{tblhead}
		\hd{Module} & \hd{Specification} & \hd{Output} \\
		Input features & Raw I/Q, half-symbol sum and difference, half-symbol indicator & $32\times(6N_{\mathrm{rx}}+1)$ \\
		Input embedding & Linear projection $+$ positional encoding, $d=192$ & $32\times d$ \\
		Waveform encoder & $L_e=6$, $n_h=8$, $d_{\mathrm{ff}}=384$, dropout $0.1$ & $32\times d$ \\
		Candidate vectors & $M=4$ trainable vectors & $M\times d$ \\
		Candidate extraction & $L_c=2$, $n_h=8$, $d_{\mathrm{ff}}=384$, dropout $0.1$ & $M\times d$ \\
		Presence head & MLP with hidden dimension $384$ & $M\times 1$ \\
		Channel head & MLP with hidden dimension $384$ & $M\times 2N_{\mathrm{rx}}$ \\
		Bit decoder & $L_b=2$, $n_h=8$, $d_{\mathrm{ff}}=384$, dropout $0.1$ & $M\times 16$ \\
	\end{tabularx}
	\vspace{-0.5cm}
\end{table}

\subsection{Training and Inference}
\label{subsec:satr_training}
SATR is trained offline to recover unordered RN16 sequences from collided slots generated by the baseband model in Section~\ref{sec:system_model}. For a slot with $R\leq M$ responding tags, the true RN16 sequences form an unordered set, while SATR produces $M$ candidate outputs. We therefore construct $M$ training targets: $R$ targets correspond to the responding tags, and the remaining $M-R$ targets represent absent tags. The presence indicator of target $j$ is denoted by $y_j\in\{0,1\}$. For $y_j=1$, target $j$ contains the effective channel $\mathbf{g}_j$ and the RN16 vector $\mathbf{b}_j=[b_j[0],\ldots,b_j[15]]$. For $y_j=0$, it contains only the presence indicator and has no channel or bit sequence. Because the candidate outputs and the training targets are both unordered, an assignment step determines their pairing.

For candidate output $i$ and target $j$, the assignment cost is based on the presence classification loss for target $j$ and, when the target corresponds to a responding tag, the squared error between the auxiliary channel estimate and the target channel. The assignment cost is defined as
\begin{equation}
	\mathcal{C}_{ij}
	=
	\lambda_p\,\operatorname{BCE}(\hat{p}_i,y_j)
	+
	\lambda_c\,y_j\|\hat{\mathbf{g}}_i-\mathbf{g}_j\|^2,
	\label{eq:match_cost}
\end{equation}
where $\operatorname{BCE}(\hat{p}_i,y_j)=-y_j\log \hat{p}_i-(1-y_j)\log(1-\hat{p}_i)$ is the binary cross entropy. The nonnegative weights $\lambda_p$ and $\lambda_c$ control the relative contributions of the presence and channel costs. The multiplier $y_j$ disables the channel cost for absent targets, so they contribute only the presence classification loss to the assignment cost.

The candidate outputs are assigned to the targets by minimizing the total assignment cost over all one-to-one pairings. The optimal assignment is given by
\begin{equation}
	\pi^\star
	=
	\argmin_{\pi\in\Pi_M}
	\sum_{i=1}^{M}\mathcal{C}_{i,\pi(i)},
	\label{eq:hungarian}
\end{equation}
where $\Pi_M$ denotes the set of all permutations of the target indices $\{1,\ldots,M\}$. Each permutation $\pi\in\Pi_M$ specifies one possible assignment, in which candidate output $i$ is assigned to target $\pi(i)$. The optimal assignment $\pi^\star$ is then used to select the target labels for each candidate output. The bit loss is excluded from the assignment cost and evaluated only after the assignment, when the RN16 label is known for each candidate assigned to a responding tag.

After the assignment is obtained, each candidate output inherits the presence label of its assigned target, i.e.,
\begin{equation}
	y_i^{\mathrm{m}} = y_{\pi^\star(i)}, \qquad i=1,\ldots,M.
\end{equation}
The candidate outputs assigned to responding tags form the active set, i.e., $\mathcal{I}_{\mathrm{act}}=\{ i\in\{1,\ldots,M\} \mid y_i^{\mathrm{m}}=1 \}$. Only these active candidates have channel and RN16 labels, which are defined as
\begin{equation}
	\mathbf{g}_i^{\mathrm{m}} = \mathbf{g}_{\pi^\star(i)}, \qquad
	\mathbf{b}_i^{\mathrm{m}} = \mathbf{b}_{\pi^\star(i)}, \qquad i\in\mathcal{I}_{\mathrm{act}}.
\end{equation}

The presence loss penalizes incorrect active or absent decisions for the candidate outputs. Since each candidate output receives a presence label through the assignment, this loss is evaluated over all $M$ outputs and is given by
\begin{equation}
	\mathcal{L}_{\mathrm{pres}}
	=
	\frac{1}{M}
	\sum_{i=1}^{M}
	\operatorname{BCE}(\hat{p}_i,y_i^{\mathrm{m}}).
	\label{eq:loss_pres}
\end{equation}

The channel loss is evaluated only over active candidate outputs, for which an assigned effective channel label is available. It measures the squared error between the auxiliary effective channel estimate $\hat{\mathbf{g}}_i$ and the assigned channel label $\mathbf{g}_i^{\mathrm{m}}$, and is given by
\begin{equation}
	\mathcal{L}_{\mathrm{chan}}
	=
	\frac{1}{|\mathcal{I}_{\mathrm{act}}|}
	\sum_{i\in\mathcal{I}_{\mathrm{act}}}
	\|\hat{\mathbf{g}}_i-\mathbf{g}_i^{\mathrm{m}}\|^2.
	\label{eq:loss_chan}
\end{equation}

The bit loss is evaluated only for the active candidate outputs, for which assigned RN16 labels are available. It averages the binary cross entropy over the 16 bits and the active candidate outputs, and is given by
\begin{equation}
	\mathcal{L}_{\mathrm{bit}}
	=
	\frac{1}{16|\mathcal{I}_{\mathrm{act}}|}
	\sum_{i\in\mathcal{I}_{\mathrm{act}}}
	\sum_{k=0}^{15}
	\operatorname{BCE}(\hat{q}_i[k],b_i^{\mathrm{m}}[k]),
	\label{eq:loss_bit}
\end{equation}
where $b_i^{\mathrm{m}}[k]$ is the $k$-th bit of the assigned RN16 label for candidate output $i$.

The total training objective is a weighted sum of the presence, channel, and bit losses and is given by
\begin{equation}
	\mathcal{L}
	=
	\lambda_p\mathcal{L}_{\mathrm{pres}}
	+
	\lambda_c\mathcal{L}_{\mathrm{chan}}
	+
	\mathcal{L}_{\mathrm{bit}},
	\label{eq:loss_total}
\end{equation}
where $\lambda_p$ and $\lambda_c$ are the presence and channel weights introduced in the assignment cost in \eqref{eq:match_cost}.

After offline training, SATR maps the received samples from a collided slot to $M$ candidate outputs. Candidate output $i$ contains the presence probability $\hat{p}_i$, the auxiliary channel estimate $\hat{\mathbf{g}}_i$, and the RN16 bit probability vector $\hat{\mathbf{q}}_i=[\hat{q}_i[0],\ldots,\hat{q}_i[15]]\in(0,1)^{16}$.

At inference, candidate output $i$ is declared active if its presence probability exceeds the threshold $p_{\mathrm{th}}$. The active output indices are defined as $\hat{\mathcal{I}}=\{i\in\{1,\dots,M\}\mid \hat{p}_i > p_{\mathrm{th}}\}$, where $p_{\mathrm{th}}$ is chosen on validation data after training. The estimated collision size is the cardinality of this set given by
\begin{equation}
	\hat{R}
	=
	|\hat{\mathcal{I}}|
	=
	\sum_{i=1}^{M} \mathbbm{1}\{\hat{p}_i > p_{\mathrm{th}}\}.
	\label{eq:rhat}
\end{equation}
For each active candidate output $i\in\hat{\mathcal{I}}$, the RN16 bit estimates are obtained by thresholding the bit probabilities, i.e.,
\begin{equation}
	\hat b_i[k]
	=
	\mathbbm{1}\{\hat q_i[k]>1/2\},
	\qquad k=0,\ldots,15.
\end{equation}
The decoded RN16 estimate for candidate output $i$ is $\hat{\mathbf{b}}_i=[\hat b_i[0],\ldots,\hat b_i[15]]$, and SATR returns the unordered set $\{\hat{\mathbf{b}}_i:i\in\hat{\mathcal{I}}\}$. Thus, at inference, SATR estimates the number of responding tags and decodes their RN16 sequences directly from the received samples, while the auxiliary channel estimate is used only during training for assignment and channel supervision.

For offline training, synthetic collision waveforms are generated from the baseband model with up to four responding tags. The training dataset covers random RN16 sequences, independent effective channel realizations, and SNRs uniformly sampled from $0$ to $30$~dB. For each generated slot, the responding tag indicators, RN16 sequences, and effective channels are known and used as labels for the assignment and loss computation. The main data generation, optimization, and threshold selection settings are summarized in Table~\ref{tab:satr_training_settings}.

\begin{table}[t]
	\centering
	\caption{SATR training settings}
	\label{tab:satr_training_settings}
	\footnotesize
	\setlength{\tabcolsep}{3pt}
	\renewcommand{\arraystretch}{1.25}
	\rowcolors{2}{white}{tblzebra}
	\begin{tabularx}{0.95\columnwidth}{@{}>{\raggedright\arraybackslash}p{0.30\columnwidth}>{\raggedright\arraybackslash}X@{}}
		\rowcolor{tblhead}
		\hd{Parameter} & \hd{Value} \\
		Candidate outputs & $M=4$ \\
		Receive antennas & $N_{\mathrm{rx}}=1$ \\
		Collision size PMF & $\Pr(R=1)=0.15$, $\Pr(R=2)=0.20$, $\Pr(R=3)=0.25$, $\Pr(R=4)=0.40$ \\
		RN16 vectors & $\mathbf{b}_i\sim\mathrm{Unif}(\{0,1\}^{16})$ \\
		Effective channels & $\mathbf{g}_i\overset{\mathrm{i.i.d.}}{\sim}\CN(\mathbf{0},\mathbf{I}_{N_{\mathrm{rx}}})$ \\
		Training SNR & $\mathrm{SNR}_{\mathrm{dB}}\sim\mathcal{U}(0,30)$ \\
		Batch size & $256$ \\
		Optimizer & AdamW \\
		Training steps & $5\times10^{5}$ \\
		Learning rate & $5\times10^{-4}$ after $2000$ warmup steps, cosine decay to $5\times10^{-5}$ \\
		Weight decay & $10^{-5}$ \\
		Gradient clipping & Global norm clipped to $1.0$ \\
		Loss weights & $\lambda_c=0.05$, $\lambda_p=1.0$ \\
		Presence threshold & $p_{\mathrm{th}}=0.70$, selected on a validation dataset \\
	\end{tabularx}
	\vspace{-0.6cm}
\end{table}

\section{Numerical Results}
\label{sec:results}

We now represent the numerical results that evaluate the collision recovery performance of SATR and the resulting FSA throughput. After describing the simulation setup, we assess the decoding performance of SATR using tag count estimation accuracy and tag success rate across collision sizes and SNR values. We then evaluate the FSA throughput achieved by SATR for different $(M,J)$ recovery configurations. Finally, we compare SATR with a BCJR detector with perfect CSI~\cite{bahl1974optimal} and the collision recovery method of Alfayoumi et al.~\cite{alfayoumi2025nextgen} for the recovery configuration $M=4$ under single acknowledgment and full recovery.

In the simulations, collided slots are generated using the baseband collision model in \eqref{eq:baseband_signal} with a single receive antenna, $N_{\mathrm{rx}}=1$. For each collision size $R\in\{1,2,3,4\}$, a slot contains $R$ responding tags, and each tag is assigned an independent RN16 sequence encoded with FM0.\footnote{The simulations use the centered bipolar FM0 representation with half-symbol levels in $\{-1,+1\}$. This representation preserves the RN16 sequence and FM0 transitions and contains no carrier leakage component.} We consider a flat-fading channel model where channel coefficients are fixed over the RN16 response for each tag. The effective channel of tag $i$ is a scalar coefficient $g_i$, drawn independently as $g_i\sim\CN(0,1)$. The received baseband waveform is sampled once per FM0 half symbol as in \eqref{eq:star_sampling}, producing $32$ complex samples per slot corrupted by AWGN. All RN16 responses, channel coefficients, and noise samples are regenerated for every slot realization. Each collision size is evaluated from $0$ to $30$ dB in $5$ dB steps using $10^{6}$ tag responses per SNR point.

All SATR results are obtained with a single model trained offline using the settings summarized in Table~\ref{tab:satr_training_settings}. For every evaluated slot, the trained model estimates the collision size and decodes the RN16 sequences from the received samples alone, with the presence threshold fixed to $p_{\mathrm{th}}=0.70$ for all collision sizes and SNR values. No retraining or parameter adaptation is performed across operating points.

\begin{figure}[t]
	\centering
	\begin{tikzpicture}
		\pgfplotsset{
			satrcurve/.style={line width=1.0pt, mark size=1.7pt},
			satrRone/.style={satrcurve, color=figcolor2, mark=*},
			satrRtwo/.style={satrcurve, color=figcolor5, mark=square*},
			satrRthree/.style={satrcurve, color=figcolor3, mark=triangle*},
			satrRfour/.style={satrcurve, color=figcolor1, mark=diamond*},
		}
		\begin{axis}[
			width=\columnwidth,
			height=0.65\columnwidth,
			xmin=0, xmax=30,
			ymin=0, ymax=1.03,
			xtick={0,5,10,15,20,25,30},
			ytick={0,0.2,0.4,0.6,0.8,1.0},
			xlabel={SNR (dB)},
			ylabel={Tag count estimation accuracy},
			grid=major,
			grid style={dashed, gray!25},
			tick label style={font=\scriptsize},
			label style={font=\footnotesize},
			legend style={
				at={(0.5,-0.22)},
				anchor=north,
				legend columns=4,
				font=\scriptsize,
				draw=none,
				/tikz/every even column/.append style={column sep=5pt}
			},
		]
			\addplot[satrRone] coordinates {(0,0.7162) (5,0.9522) (10,0.9861) (15,0.9957) (20,0.9984) (25,0.9994) (30,1.0000)};
			\addlegendentry{$R=1$}

			\addplot[satrRtwo] coordinates {(0,0.4796) (5,0.7361) (10,0.8933) (15,0.9598) (20,0.9850) (25,0.9941) (30,0.9982)};
			\addlegendentry{$R=2$}

			\addplot[satrRthree] coordinates {(0,0.4953) (5,0.6325) (10,0.8275) (15,0.9312) (20,0.9713) (25,0.9904) (30,0.9970)};
			\addlegendentry{$R=3$}

			\addplot[satrRfour] coordinates {(0,0.5746) (5,0.6387) (10,0.8087) (15,0.9206) (20,0.9699) (25,0.9887) (30,0.9935)};
			\addlegendentry{$R=4$}
		\end{axis}
	\end{tikzpicture}
	\caption{Tag count estimation accuracy of SATR for collision sizes $R\in\{1,2,3,4\}$.}
	\label{fig:satr_tag_count_accuracy}
	\vspace{-0.5cm}
\end{figure}
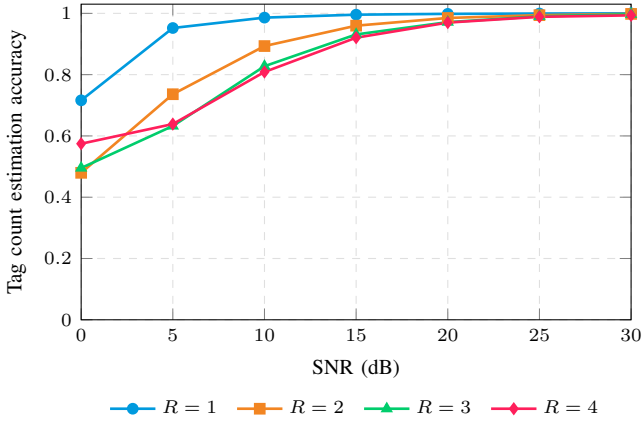

Fig.~\ref{fig:satr_tag_count_accuracy} shows the tag count estimation accuracy of SATR, defined as the probability that the estimated number of responding tags equals the true collision size. At low SNR, noise reduces the separation between the received I/Q patterns and weakens the FM0 transition structure, so the estimated count is often confused with nearby collision sizes. As the SNR increases, these waveform features become clearer, and the accuracy improves for all collision sizes. The single tag case is identified most reliably, while larger collisions are more difficult because additional overlapping responses produce more complex received waveforms. At high SNR, the tag count estimation accuracy approaches one for all tested collision sizes, showing that SATR reliably estimates the number of responding tags from the received waveform alone.

\begin{figure}[t]
	\centering
	\begin{tikzpicture}
		\pgfplotsset{
			satrcurve/.style={line width=1.0pt, mark size=1.7pt},
			bcjrcurve/.style={line width=0.8pt, dash pattern=on 2.4pt off 1.6pt, mark size=1.5pt, mark options={solid}},
			legend image code/.code={
				\draw[mark repeat=2, mark phase=2] plot coordinates {(0cm,0cm) (0.25cm,0cm) (0.5cm,0cm)};
			},
		}
		\begin{axis}[
			width=\columnwidth,
			height=0.65\columnwidth,
			xmin=0, xmax=30,
			ymin=0, ymax=1.03,
			xtick={0,5,10,15,20,25,30},
			ytick={0,0.2,0.4,0.6,0.8,1.0},
			xlabel={SNR (dB)},
			ylabel={Tag success rate},
			grid=major,
			grid style={dashed, gray!25},
			tick label style={font=\scriptsize},
			label style={font=\footnotesize},
			legend style={
				at={(0.5,-0.20)},
				anchor=north,
				legend columns=4,
				font=\tiny,
				draw=none,
				/tikz/every even column/.append style={column sep=3pt}
			},
			legend cell align={left},
		]
			\addplot[satrcurve, color=figcolor2, mark=*] coordinates {(0,0.4207) (5,0.7339) (10,0.9035) (15,0.9681) (20,0.9896) (25,0.9967) (30,0.9989)};
			\addlegendentry{SATR $R=1$}
			\addplot[satrcurve, color=figcolor5, mark=square*] coordinates {(0,0.2981) (5,0.6050) (10,0.8358) (15,0.9415) (20,0.9802) (25,0.9928) (30,0.9970)};
			\addlegendentry{SATR $R=2$}
			\addplot[satrcurve, color=figcolor3, mark=triangle*] coordinates {(0,0.1889) (5,0.4328) (10,0.7124) (15,0.8846) (20,0.9551) (25,0.9798) (30,0.9883)};
			\addlegendentry{SATR $R=3$}
			\addplot[satrcurve, color=figcolor1, mark=diamond*] coordinates {(0,0.1123) (5,0.2516) (10,0.4830) (15,0.7142) (20,0.8457) (25,0.8984) (30,0.9152)};
			\addlegendentry{SATR $R=4$}
			\addplot[bcjrcurve, color=figcolor2, mark=*] coordinates {(0,0.4572) (5,0.7598) (10,0.9133) (15,0.9713) (20,0.9908) (25,0.9972) (30,0.9990)};
			\addlegendentry{BCJR $R=1$}
			\addplot[bcjrcurve, color=figcolor5, mark=square*] coordinates {(0,0.3461) (5,0.6554) (10,0.8624) (15,0.9527) (20,0.9846) (25,0.9951) (30,0.9984)};
			\addlegendentry{BCJR $R=2$}
			\addplot[bcjrcurve, color=figcolor3, mark=triangle*] coordinates {(0,0.2399) (5,0.5230) (10,0.7823) (15,0.9204) (20,0.9735) (25,0.9914) (30,0.9971)};
			\addlegendentry{BCJR $R=3$}
			\addplot[bcjrcurve, color=figcolor1, mark=diamond*] coordinates {(0,0.1547) (5,0.3755) (10,0.6652) (15,0.8651) (20,0.9532) (25,0.9849) (30,0.9952)};
			\addlegendentry{BCJR $R=4$}
		\end{axis}
	\end{tikzpicture}
	\caption{Tag success rate of SATR (solid) and the BCJR detector with perfect CSI (dashed) for collision sizes $R\in\{1,2,3,4\}$.}
	\label{fig:satr_tag_success}
	\vspace{-0.5cm}
\end{figure}
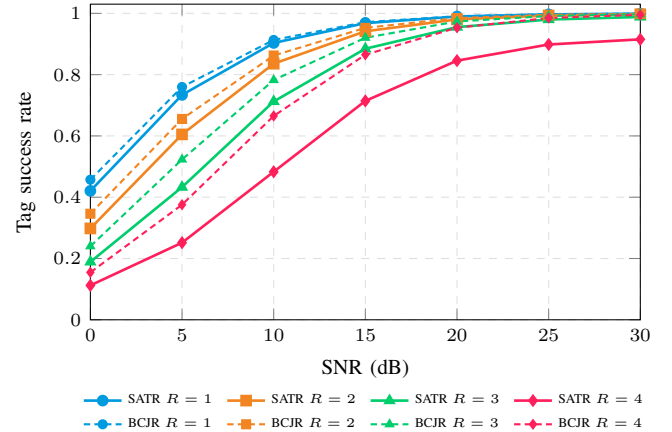

Fig.~\ref{fig:satr_tag_success} depicts a comparison of the tag success rates achieved by SATR and the BCJR detector with perfect CSI.\footnote{The BCJR detector with perfect CSI is used as an informed reference. It is given the true collision size, the effective channels, and the noise variance, and performs maximum a posteriori decoding over the joint FM0 trellis.} A tag is counted as successful only when all $16$ bits of its RN16 sequence are recovered correctly under a one-to-one matching between the decoded and transmitted sequences. The success rate increases with SNR and decreases as the collision size grows, reflecting the difficulty of separating multiple overlapping RN16 responses. For up to three colliding tags, SATR remains close to the BCJR detector without using the true collision size or channel coefficients, and the gap becomes almost negligible at high SNR. For $R=4$, the gap remains more visible, making it the most challenging case for SATR.

We next evaluate how SATR improves the FSA throughput by recovering responses from collided slots. Each FSA frame contains $K=1000$ slots, and each tag independently chooses one slot uniformly. For each recovery configuration $(M,J)$ with $M\leq4$, the tag population $N$ is selected to maximize the ideal finite-frame throughput. This places each configuration at its analytically optimal operating point, assuming that collisions of up to $M$ tags are recoverable and at most $J$ decoded responses can be acknowledged in each slot. SATR is then applied to the slots within the recoverable range, while slots with more than $M$ responding tags are treated as unresolved. This matches the recovery model, which counts only collisions of up to $M$ tags.\footnote{For slots with five or six responding tags, SATR still returns up to four candidate responses and often recovers some of them correctly. Excluding these slots therefore slightly underestimates the achievable throughput.} If $D$ responses are recovered from a slot, the slot contributes $\min(D,J)$ acknowledged tags. The reported throughput is the average number of acknowledged tags per slot over $100$ independent frames.

\begin{figure}[t]
	\centering
	\begin{tikzpicture}
		\pgfplotsset{
			satrcurve/.style={line width=1.0pt, mark size=1.7pt},
			idealline/.style={line width=0.6pt, dotted},
			legend image code/.code={
				\draw[mark repeat=2, mark phase=2] plot coordinates {(0cm,0cm) (0.25cm,0cm) (0.5cm,0cm)};
			},
		}
		\begin{axis}[
			width=\columnwidth,
			height=0.65\columnwidth,
			xmin=0, xmax=30,
			ymin=0, ymax=0.9,
			xtick={0,5,10,15,20,25,30},
			ytick={0,0.2,0.4,0.6,0.8},
			xlabel={SNR (dB)},
			ylabel={Expected throughput},
			grid=major,
			grid style={dashed, gray!25},
			tick label style={font=\scriptsize},
			label style={font=\footnotesize},
			legend style={
				at={(0.5,-0.20)},
				anchor=north,
				legend columns=4,
				font=\scriptsize,
				draw=none,
				/tikz/every even column/.append style={column sep=5pt}
			},
			legend cell align={left},
		]
			\addplot[idealline, forget plot, color=figcolor2] coordinates {(0,0.3681)(30,0.3681)};
			\addplot[idealline, forget plot, color=figcolor5] coordinates {(0,0.5872)(30,0.5872)};
			\addplot[idealline, forget plot, color=figcolor3] coordinates {(0,0.7263)(30,0.7263)};
			\addplot[idealline, forget plot, color=figcolor1] coordinates {(0,0.8169)(30,0.8169)};
			\addplot[satrcurve, color=figcolor2, mark=*] coordinates {(0,0.1537) (5,0.2710) (10,0.3338) (15,0.3606) (20,0.3640) (25,0.3690) (30,0.3684)};
			\addlegendentry{$M=1$}
			\addplot[satrcurve, color=figcolor5, mark=square*] coordinates {(0,0.2633) (5,0.4475) (10,0.5377) (15,0.5689) (20,0.5828) (25,0.5879) (30,0.5865)};
			\addlegendentry{$M=2$}
			\addplot[satrcurve, color=figcolor3, mark=triangle*] coordinates {(0,0.3296) (5,0.5546) (10,0.6637) (15,0.7045) (20,0.7194) (25,0.7244) (30,0.7241)};
			\addlegendentry{$M=3$}
			\addplot[satrcurve, color=figcolor1, mark=diamond*] coordinates {(0,0.3632) (5,0.6077) (10,0.7417) (15,0.7933) (20,0.8073) (25,0.8149) (30,0.8147)};
			\addlegendentry{$M=4$}
		\end{axis}
	\end{tikzpicture}
	\caption{Expected throughput of SATR with single acknowledgment ($J=1$) for recovery configurations $M\in\{1,\ldots,4\}$. The dotted lines mark the ideal theoretical limit for each configuration.}
	\label{fig:satr_throughput_j1}
	\vspace{-0.5cm}
\end{figure}

Fig.~\ref{fig:satr_throughput_j1} shows the FSA throughput of SATR under the single acknowledgment setting, $J=1$. In this setting, a recovered collision slot contributes one acknowledged tag, even if more than one response is decoded from that slot. The configuration $M=1$ reduces to conventional FSA, since only singleton slots are recovered. As the SNR increases, more collision slots are successfully recovered, and the throughput for each value of $M$ approaches its corresponding ideal limit. Increasing $M$ raises this limit because collisions involving more tags can contribute to the throughput. For $M=4$, SATR reaches about $0.815$ tags per slot, more than twice the conventional FSA limit of $1/e$.

\begin{figure}[t]
	\centering
	\begin{tikzpicture}
		\pgfplotsset{
			satrcurve/.style={line width=1.0pt, mark size=1.7pt},
			idealline/.style={line width=0.6pt, dotted},
			legend image code/.code={
				\draw[mark repeat=2, mark phase=2] plot coordinates {(0cm,0cm) (0.25cm,0cm) (0.5cm,0cm)};
			},
		}
		\begin{axis}[
			width=\columnwidth,
			height=0.65\columnwidth,
			xmin=0, xmax=30,
			ymin=0, ymax=2.1,
			xtick={0,5,10,15,20,25,30},
			ytick={0,0.4,0.8,1.2,1.6,2.0},
			xlabel={SNR (dB)},
			ylabel={Expected throughput},
			grid=major,
			grid style={dashed, gray!25},
			tick label style={font=\scriptsize},
			label style={font=\footnotesize},
			legend style={
				at={(0.5,-0.20)},
				anchor=north,
				legend columns=3,
				font=\scriptsize,
				draw=none,
				/tikz/every even column/.append style={column sep=4pt}
			},
			legend cell align={left},
		]
			\addplot[idealline, forget plot, color=figcolor2] coordinates {(0,0.8403)(30,0.8403)};
			\addplot[idealline, forget plot, color=figcolor4] coordinates {(0,1.1757)(30,1.1757)};
			\addplot[idealline, forget plot, color=figcolor5] coordinates {(0,1.3716)(30,1.3716)};
			\addplot[idealline, forget plot, color=figcolor6] coordinates {(0,1.7823)(30,1.7823)};
			\addplot[idealline, forget plot, color=figcolor3] coordinates {(0,1.4140)(30,1.4140)};
			\addplot[idealline, forget plot, color=figcolor1] coordinates {(0,1.9431)(30,1.9431)};
			\addplot[satrcurve, color=figcolor2, mark=*] coordinates {(0,0.2891) (5,0.5505) (10,0.7250) (15,0.8027) (20,0.8232) (25,0.8374) (30,0.8375)};
			\addlegendentry{$M=2$, $J=2$}
			\addplot[satrcurve, color=figcolor4, mark=pentagon*] coordinates {(0,0.3730) (5,0.7335) (10,0.9979) (15,1.1116) (20,1.1536) (25,1.1671) (30,1.1730)};
			\addlegendentry{$M=3$, $J=2$}
			\addplot[satrcurve, color=figcolor5, mark=square*] coordinates {(0,0.3693) (5,0.7539) (10,1.0895) (15,1.2655) (20,1.3283) (25,1.3625) (30,1.3612)};
			\addlegendentry{$M=3$, $J=3$}
			\addplot[satrcurve, color=figcolor3, mark=triangle*] coordinates {(0,0.4148) (5,0.8265) (10,1.1572) (15,1.3196) (20,1.3797) (25,1.4034) (30,1.4050)};
			\addlegendentry{$M=4$, $J=2$}
			\addplot[satrcurve, color=figcolor6, mark=star] coordinates {(0,0.4056) (5,0.8567) (10,1.3066) (15,1.5784) (20,1.7025) (25,1.7488) (30,1.7571)};
			\addlegendentry{$M=4$, $J=3$}
			\addplot[satrcurve, color=figcolor1, mark=diamond*] coordinates {(0,0.4018) (5,0.8468) (10,1.3197) (15,1.6437) (20,1.7990) (25,1.8533) (30,1.8736)};
			\addlegendentry{$M=4$, $J=4$}
		\end{axis}
	\end{tikzpicture}
	\caption{Expected throughput of SATR with multiple acknowledgments for recovery configurations $(M,J)$. The dotted lines mark the ideal theoretical limit for each configuration.}
	\label{fig:satr_throughput_multi}
	\vspace{-0.5cm}
\end{figure}
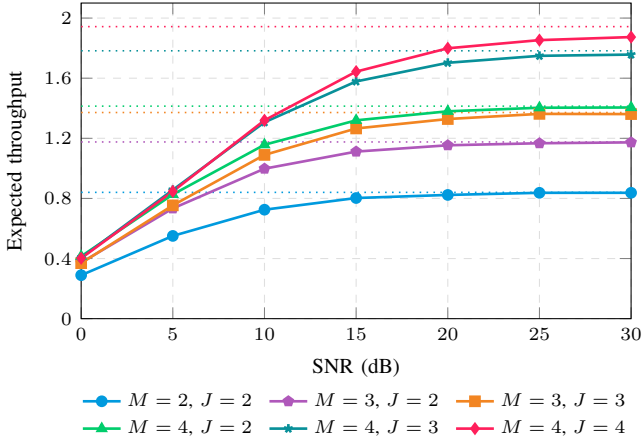

Fig.~\ref{fig:satr_throughput_multi} shows the FSA throughput of SATR when multiple responses can be acknowledged from the same recovered slot. Unlike the single acknowledgment setting, the throughput now depends on how many responses SATR recovers and the reader can acknowledge from each collision. As the SNR increases, the throughput improves for all recovery configurations and approaches the corresponding ideal limit. The remaining gap is most visible for $M=4$ with $J=4$, since reaching its ideal limit requires recovering all four responses from a four tag collision. For a fixed acknowledgment limit, increasing $M$ improves throughput by making larger collisions recoverable, while, for a fixed $M$, increasing $J$ allows more recovered responses from each slot to contribute. For the full recovery configuration, $M=4$ with $J=4$, SATR reaches $1.87$ tags per slot, more than five times the conventional FSA limit.

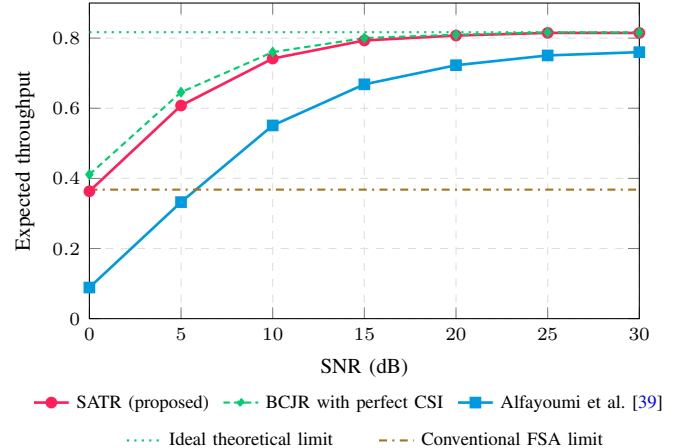
\begin{figure}[t]
	\centering
	\begin{tikzpicture}
		\pgfplotsset{
			methodcurve/.style={line width=1.0pt, mark size=1.7pt},
			bcjrcurve/.style={line width=0.8pt, dash pattern=on 2.4pt off 1.6pt, mark size=1.5pt, mark options={solid}},
			idealline/.style={line width=0.8pt, dotted},
			convline/.style={line width=0.8pt, dash dot},
		}
		\begin{axis}[
			width=\columnwidth,
			height=0.65\columnwidth,
			xmin=0, xmax=30,
			ymin=0, ymax=0.9,
			xtick={0,5,10,15,20,25,30},
			ytick={0,0.2,0.4,0.6,0.8},
			xlabel={SNR (dB)},
			ylabel={Expected throughput},
			clip=false,
			grid=major,
			grid style={dashed, gray!25},
			tick label style={font=\scriptsize},
			label style={font=\footnotesize},
		]
			\addplot[methodcurve, color=figcolor1, mark=*] coordinates {(0,0.3632) (5,0.6077) (10,0.7417) (15,0.7933) (20,0.8073) (25,0.8149) (30,0.8147)};
			\addplot[bcjrcurve, color=figcolor3, mark=diamond*] coordinates {(0,0.4109) (5,0.6458) (10,0.7600) (15,0.8002) (20,0.8097) (25,0.8165) (30,0.8157)};
			\addplot[methodcurve, color=figcolor2, mark=square*] coordinates {(0,0.0887) (5,0.3323) (10,0.5509) (15,0.6682) (20,0.7227) (25,0.7506) (30,0.7599)};
			\addplot[idealline, color=figcolor3] coordinates {(0,0.8169)(30,0.8169)};
			\addplot[convline, color=convcol] coordinates {(0,0.3679)(30,0.3679)};
			\node[anchor=north, inner sep=0pt, font=\scriptsize] at (axis description cs:0.47,-0.24) {%
				\tikz[baseline=-0.6ex]{\draw[line width=1.0pt, color=figcolor1, mark repeat=2, mark phase=2] plot[mark=*, mark size=1.7pt] coordinates {(0cm,0cm) (0.23cm,0cm) (0.46cm,0cm)};}~SATR (proposed)%
				\hspace{5pt}%
				\tikz[baseline=-0.6ex]{\draw[line width=0.8pt, dash pattern=on 2.4pt off 1.6pt, color=figcolor3, mark repeat=2, mark phase=2] plot[mark=diamond*, mark size=1.5pt] coordinates {(0cm,0cm) (0.23cm,0cm) (0.46cm,0cm)};}~BCJR with perfect CSI%
				\hspace{5pt}%
				\tikz[baseline=-0.6ex]{\draw[line width=1.0pt, color=figcolor2, mark repeat=2, mark phase=2] plot[mark=square*, mark size=1.7pt] coordinates {(0cm,0cm) (0.23cm,0cm) (0.46cm,0cm)};}~Alfayoumi et al.~\cite{alfayoumi2025nextgen}%
			};
			\node[anchor=north, inner sep=0pt, font=\scriptsize] at (axis description cs:0.5,-0.36) {%
				\tikz[baseline=-0.6ex]{\draw[line width=0.8pt, dotted, color=figcolor3] (0cm,0cm) -- (0.46cm,0cm);}~Ideal theoretical limit
				\hspace{12pt}
				\tikz[baseline=-0.6ex]{\draw[line width=0.8pt, dash dot, color=convcol] (0cm,0cm) -- (0.46cm,0cm);}~Conventional FSA limit%
			};
		\end{axis}
	\end{tikzpicture}
	\caption{Expected throughput comparison for $M=4$ with single acknowledgment ($J=1$).}
	\label{fig:throughput_comparison_j1}
	\vspace{-0.5cm}
\end{figure}

Fig.~\ref{fig:throughput_comparison_j1} depicts the throughput comparison among SATR, the BCJR detector with perfect CSI, and the I/Q geometry method of Alfayoumi et al.~\cite{alfayoumi2025nextgen} for $M=4$ under single acknowledgment. SATR closely follows the BCJR reference over the SNR range and approaches the ideal limit at high SNR. The method of Alfayoumi et al.~\cite{alfayoumi2025nextgen} also improves with SNR but gradually levels off below SATR and the BCJR reference. Its constellation labeling relies on observing a complete and well separated set of load-state clusters. As the collision size increases, a finite RN16 waveform is less likely to contain every load-state combination, consistent with the cluster observability analysis in \cite{alfayoumi2025nextgen}. As a result, some slots remain unresolved even as the SNR increases.

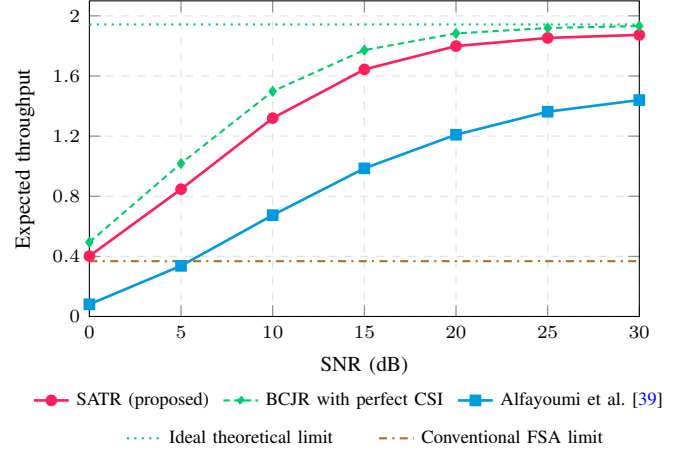
\begin{figure}[t]
	\centering
	\begin{tikzpicture}
		\pgfplotsset{
			methodcurve/.style={line width=1.0pt, mark size=1.7pt},
			bcjrcurve/.style={line width=0.8pt, dash pattern=on 2.4pt off 1.6pt, mark size=1.5pt, mark options={solid}},
			idealline/.style={line width=0.8pt, dotted},
			convline/.style={line width=0.8pt, dash dot},
		}
		\begin{axis}[
			width=\columnwidth,
			height=0.65\columnwidth,
			xmin=0, xmax=30,
			ymin=0, ymax=2.1,
			xtick={0,5,10,15,20,25,30},
			ytick={0,0.4,0.8,1.2,1.6,2.0},
			xlabel={SNR (dB)},
			ylabel={Expected throughput},
			clip=false,
			grid=major,
			grid style={dashed, gray!25},
			tick label style={font=\scriptsize},
			label style={font=\footnotesize},
		]
			\addplot[methodcurve, color=figcolor1, mark=*] coordinates {(0,0.4018) (5,0.8468) (10,1.3197) (15,1.6437) (20,1.7990) (25,1.8533) (30,1.8736)};
			\addplot[bcjrcurve, color=figcolor3, mark=diamond*] coordinates {(0,0.4942) (5,1.0182) (10,1.4983) (15,1.7723) (20,1.8833) (25,1.9194) (30,1.9325)};
			\addplot[methodcurve, color=figcolor2, mark=square*] coordinates {(0,0.0809) (5,0.3369) (10,0.6742) (15,0.9856) (20,1.2095) (25,1.3625) (30,1.4399)};
			\addplot[idealline, color=figcolor3] coordinates {(0,1.9431)(30,1.9431)};
			\addplot[convline, color=convcol] coordinates {(0,0.3679)(30,0.3679)};
			\node[anchor=north, inner sep=0pt, font=\scriptsize] at (axis description cs:0.47,-0.24) {%
				\tikz[baseline=-0.6ex]{\draw[line width=1.0pt, color=figcolor1, mark repeat=2, mark phase=2] plot[mark=*, mark size=1.7pt] coordinates {(0cm,0cm) (0.23cm,0cm) (0.46cm,0cm)};}~SATR (proposed)%
				\hspace{5pt}%
				\tikz[baseline=-0.6ex]{\draw[line width=0.8pt, dash pattern=on 2.4pt off 1.6pt, color=figcolor3, mark repeat=2, mark phase=2] plot[mark=diamond*, mark size=1.5pt] coordinates {(0cm,0cm) (0.23cm,0cm) (0.46cm,0cm)};}~BCJR with perfect CSI%
				\hspace{5pt}%
				\tikz[baseline=-0.6ex]{\draw[line width=1.0pt, color=figcolor2, mark repeat=2, mark phase=2] plot[mark=square*, mark size=1.7pt] coordinates {(0cm,0cm) (0.23cm,0cm) (0.46cm,0cm)};}~Alfayoumi et al.~\cite{alfayoumi2025nextgen}%
			};
			\node[anchor=north, inner sep=0pt, font=\scriptsize] at (axis description cs:0.5,-0.36) {%
				\tikz[baseline=-0.6ex]{\draw[line width=0.8pt, dotted, color=figcolor3] (0cm,0cm) -- (0.46cm,0cm);}~Ideal theoretical limit
				\hspace{12pt}
				\tikz[baseline=-0.6ex]{\draw[line width=0.8pt, dash dot, color=convcol] (0cm,0cm) -- (0.46cm,0cm);}~Conventional FSA limit%
			};
		\end{axis}
	\end{tikzpicture}
	\caption{Expected throughput comparison for $M=4$ under full recovery ($J=4$).}
	\label{fig:throughput_comparison_j4}
	\vspace{-0.7cm}
\end{figure}

Fig.~\ref{fig:throughput_comparison_j4} presents the corresponding comparison for $M=4$ under full recovery. Under $J=4$, the throughput reflects the total number of recovered responses, whereas under $J=1$, recovering any response is sufficient to acknowledge one tag. Although SATR requires no CSI, its throughput remains close to that of the BCJR detector with perfect CSI and approaches the ideal limit as SNR increases. The small remaining gap between the two methods mainly reflects the difficulty of recovering all responses in the largest collisions. The I/Q geometry method of Alfayoumi et al.~\cite{alfayoumi2025nextgen} more often recovers only a subset of responses from larger collisions, leading to a larger gap from SATR and BCJR.

We finally compare the computational cost of the three decoding methods. For a collision of size $R$, the joint FM0 trellis used by the BCJR detector contains $2^R$ load states and $4^R$ state transitions per bit, so its cost grows exponentially with $R$. The complexity of the I/Q geometry method of Alfayoumi et al.~\cite{alfayoumi2025nextgen} also scales with the collision size, since up to $2^R$ constellation clusters must be estimated and labeled before the FM0 streams are reconstructed. SATR instead applies the same fixed computation to every received slot: the input length, the number of candidate outputs ($M=4$), and the RN16 length are all independent of the realized collision size $R$.

\begin{table}[!h]
	\centering
	\caption{Mean decoding time per collision slot in milliseconds (ms)}
	\label{tab:decoding_time}
	\footnotesize
	\setlength{\tabcolsep}{3pt}
	\renewcommand{\arraystretch}{1.25}
	\rowcolors{2}{white}{tblzebra}
	\begin{tabularx}{0.95\columnwidth}{@{}>{\raggedright\arraybackslash}X*{4}{>{\centering\arraybackslash}p{0.125\columnwidth}}@{}}
		\rowcolor{tblhead}
		\hd{Method} & \hd{$R{=}1$} & \hd{$R{=}2$} & \hd{$R{=}3$} & \hd{$R{=}4$} \\
		SATR (CPU) & $14.7$ & $14.2$ & $14.1$ & $14.8$ \\
		SATR (GPU) & $0.58$ & $0.62$ & $0.61$ & $0.60$ \\
		BCJR with perfect CSI & $1.7$ & $2.4$ & $3.3$ & $4.3$ \\
		Alfayoumi et al.~\cite{alfayoumi2025nextgen} & $59.1$ & $58.9$ & $60.0$ & $60.3$ \\
	\end{tabularx}
	\vspace{-0.2 cm}
\end{table}

Table~\ref{tab:decoding_time} reports the mean decoding time per collision slot for the decoding stage alone.\footnote{Timing measurements are obtained on an Apple M4 Max chip. The CPU timings process one slot at a time, while the SATR GPU timings use batches of $256$ slots.} As expected from its fixed computational structure, the decoding time of SATR remains nearly constant across the considered collision sizes on both platforms. The CPU results provide a direct comparison with the reference methods. The BCJR detector is faster over the considered range because perfect CSI removes all estimation work and the joint trellis remains small, although its decoding time grows with the collision size. The method of Alfayoumi et al.~\cite{alfayoumi2025nextgen} requires substantially more time because constellation estimation and labeling must be performed separately for each slot. GPU execution is better suited to SATR because its self-attention and feedforward layers primarily involve matrix operations that can be evaluated more efficiently in parallel, reducing the processing time to approximately $0.6$~ms per slot.

\section{Experimental Results}
\label{sec:experimental_results}
In this section, we validate SATR using measured responses from commercial passive UHF-RFID tags collected with an SDR reader. We describe the reader setup, tag layout, and EPC Gen2 protocol parameters, and then examine the measured FM0 envelopes and I/Q samples. The decoding evaluation uses two complementary types of collisions. Direct collision captures evaluate SATR on simultaneous responses produced by multiple tags in the measurement setup. Composite collisions are formed by summing recorded single tag responses whose RN16 bits are known, providing a ground truth evaluation that connects the measured results to the numerical study.

\begin{figure}[t]
	\centering
	\includegraphics[width=\linewidth]{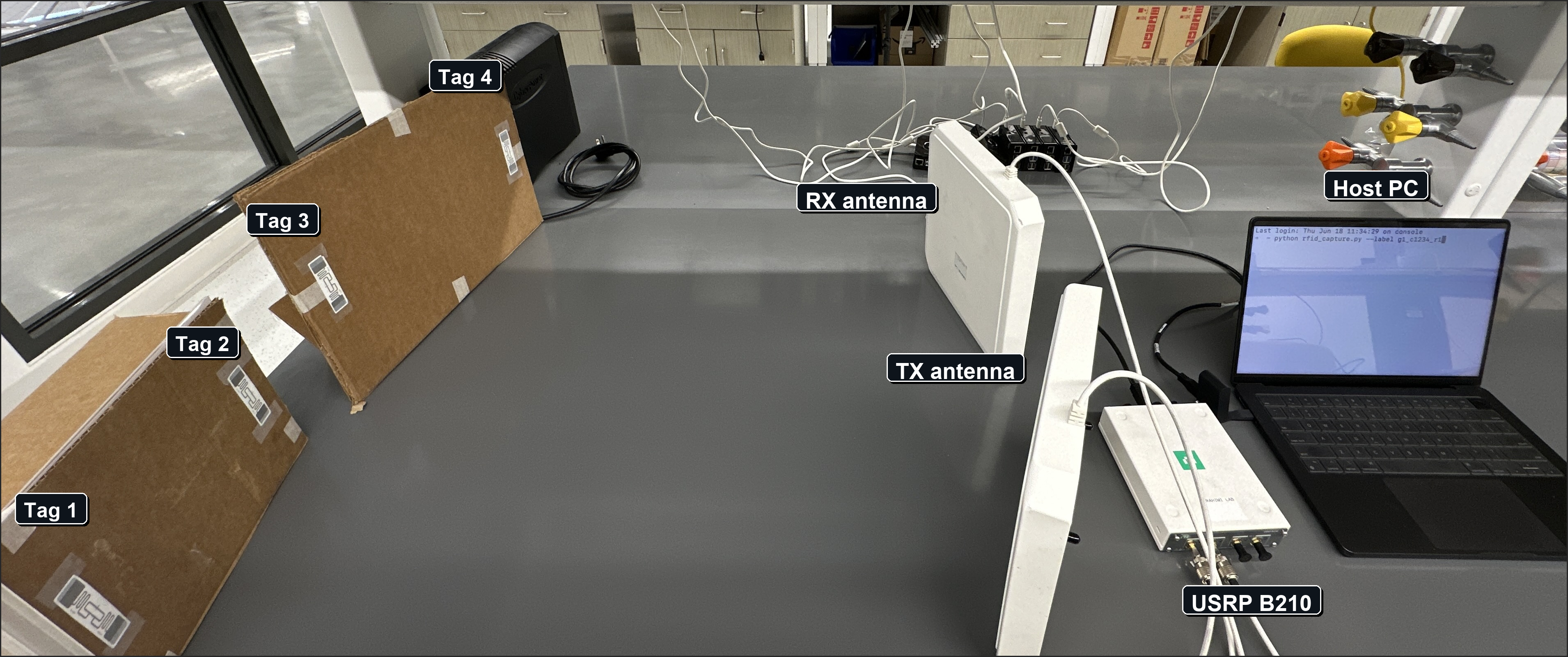}
	\caption{Experimental setup of the passive UHF-RFID reader system, based on a USRP B210 with separate transmit and receive antennas and four commercial tags at fixed positions.}
	\label{fig:measured_setup}
	\vspace{-0.2cm}
\end{figure}

The measurement setup uses a USRP B210 SDR reader implemented in GNU Radio by extending the open-source software of Kargas et al.~\cite{kargas2015fullycoherent} to record the raw I/Q samples. The reader uses a bistatic antenna configuration, with one antenna transmitting the Query command and the CW that powers the tags and a separate receive antenna capturing the backscattered response. Four commercial Alien ALN-9662 passive UHF-RFID tags are placed at fixed positions with different ranges and lateral offsets, producing distinct backscatter channels. Fig.~\ref{fig:measured_setup} shows the reader, the transmit and receive antennas, and the four tags at their labeled positions.

\FloatBarrier

All captures use the same reader setup, timing configuration, and receive processing window, so the only controlled change is the number of tags placed in the reader field. The corresponding measurement parameters are summarized in Table~\ref{tab:meas_params}. \vspace{-0.3cm}
\begin{table}[h]
	\centering
	\caption{Measurement parameters}
	\label{tab:meas_params}
	\footnotesize
	\setlength{\tabcolsep}{3pt}
	\renewcommand{\arraystretch}{1.15}
	\rowcolors{2}{white}{tblzebra}
	\begin{tabularx}{0.95\columnwidth}{@{}>{\raggedright\arraybackslash}p{0.39\columnwidth}>{\raggedright\arraybackslash}X@{}}
		\rowcolor{tblhead}
		\hd{Parameter} & \hd{Value} \\
		Center frequency & $914$~MHz \\
		Sampling rate & $2$~MS/s \\
		TX/RX gains & $86$~dB / $25$~dB \\
		Transmit amplitude & $0.7$ \\
		Tag encoding & FM0 \\
		Nominal BLF & $40$~kHz \\
		Tari & $25~\mu$s \\
		Gen2 settings & $Q=0$, $\mathrm{TRext}=0$ \\
		Response timing & $T_1=250~\mu$s, $T_2=500~\mu$s \\
		Samples per FM0 bit & $50$ \\
		Response window & $6$ bit preamble and RN16 \\
	\end{tabularx}
	\vspace{-0.6cm}
\end{table}

\begin{figure}[b]
	\centering
	\resizebox{0.9\columnwidth}{!}{\input{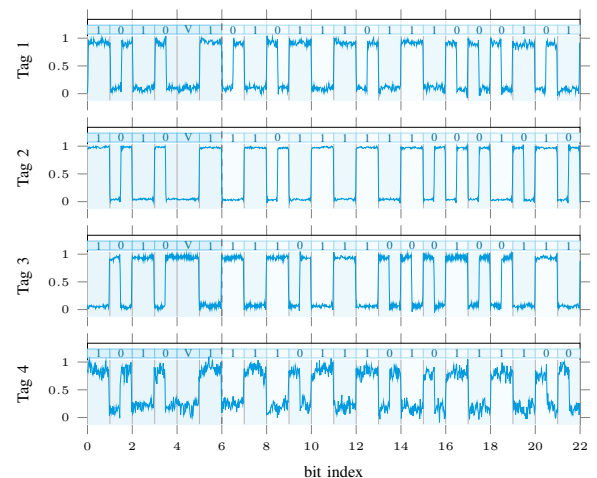}}
	\caption{Measured FM0 response envelopes for four individual tags. The first six intervals show the short preamble, followed by the RN16 bits.}
	\label{fig:single_tag_fm0_envelopes}
	\vspace{-0.5cm}
\end{figure}

\begin{figure*}[t]
	\centering
	\includegraphics[width=\textwidth]{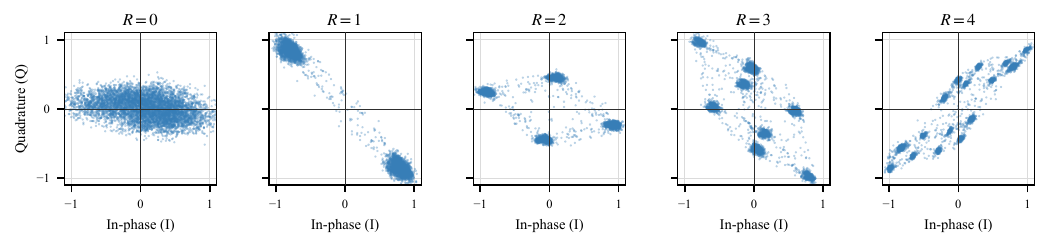}
		\caption{Measured I/Q samples for selected validation captures with no tag and one to four responding tags. All aligned response samples are plotted for each case.}
	\label{fig:measured_iq_clusters}
	\vspace{-0.5cm}
\end{figure*}

Before analyzing collisions, we inspect the single tag captures to confirm that the measured responses exhibit the expected FM0 timing and transition structure. Fig.~\ref{fig:single_tag_fm0_envelopes} shows one envelope from each tag after alignment to the FM0 preamble. The first six bit intervals correspond to the preamble, followed by the RN16 bits. The visible transitions confirm that the captured responses are consistent with FM0 encoding.

Fig.~\ref{fig:measured_iq_clusters} shows the measured I/Q samples for selected captures with no tag and with $R=1,\ldots,4$ responding tags. The no tag panel shows the receiver carrier leakage in the absence of a response. For $R=1$, the samples separate into two regions associated with the two tag load states. As $R$ increases, the I/Q pattern changes because the reader receives the combined waveform of multiple FM0 responses in the same slot. These measurements confirm that the captures contain the expected I/Q structure of multi-tag collisions.

For decoding validation, each measured response window is located from the Query timing, and only the RN16 portion is retained. The complex I/Q waveform is sampled at the two FM0 half-symbol positions of each RN16 bit, giving $32$ complex samples per attempt. Before decoding, each sequence is centered by removing its mean and scaled using the median magnitude from the single tag measurements. This keeps the measured waveform as the decoder input while placing all captures on a common amplitude scale.

A challenge in evaluating direct collision captures is that commercial tags generate RN16 values internally, so the true RN16 sequence of each responding tag is not available. We therefore use the BCJR detector as a measurement reference. The single tag captures provide a channel estimate for each tag at its fixed position. Since the tag set in each collision capture is known from the measurement setup, the BCJR detector uses the corresponding channel estimates and FM0 state evolution to infer the RN16 sequence for each responding tag. SATR is evaluated on the same measured RN16 samples, but it does not use the channel estimates and directly predicts both the number of tags and the RN16 sequences.

Table~\ref{tab:meas_validation_metrics} reports the decoding results for the direct collision captures with $R=1,2,3,4$ responding tags. The tag count estimation accuracy is computed from the known tag configuration, while the RN16 bit agreement is computed against the BCJR reference. The measured captures correspond to a high SNR operating point, with an estimated single tag SNR of approximately $27$~dB. SATR estimates the correct collision size for every measured attempt, consistent with the tag count estimation results at the corresponding SNR in Fig.~\ref{fig:satr_tag_count_accuracy}. The RN16 bit agreement remains high for $R=1$ and $R=2$, and decreases for $R=3$ and $R=4$ as more FM0 responses overlap. Since the true RN16 bits of direct collision captures are not available, this decrease reflects disagreement between SATR and the BCJR reference rather than decoding performance relative to the transmitted sequences. 

\begin{table}[h]
	\centering
	\caption{Measured decoding validation with the BCJR reference}
	\label{tab:meas_validation_metrics}
	\footnotesize
	\setlength{\tabcolsep}{3pt}
	\renewcommand{\arraystretch}{1.25}
	\rowcolors{2}{white}{tblzebra}
	\begin{tabularx}{0.95\columnwidth}{@{}>{\centering\arraybackslash}p{0.3\columnwidth}>{\centering\arraybackslash}X>{\centering\arraybackslash}p{0.23\columnwidth}@{}}
		\rowcolor{tblhead}
		\hd{Case} & \hd{Tag count estimation accuracy} & \hd{RN16 bit agreement} \\
		$R=1$ ($120$ attempts) & $100.0\%$ & $100.0\%$ \\
		$R=2$ ($140$ attempts) & $100.0\%$ & $97.2\%$ \\
		$R=3$ ($80$ attempts) & $100.0\%$ & $91.7\%$ \\
		$R=4$ ($50$ attempts) & $100.0\%$ & $79.4\%$ \\
	\end{tabularx}
	\vspace{-0.2cm}
\end{table}

To evaluate SATR with known transmitted bits on measured signals, composite collisions are constructed from recorded single tag responses whose RN16 bits are available from reliable single tag decoding. Each composite sums the aligned RN16 samples of two, three, or four different tags, retaining the measured channel responses and receiver noise while keeping the transmitted bits known. For each collision size, $360$ composites are generated by drawing random response attempts over all tag subsets, and Table~\ref{tab:meas_composite_metrics} reports the results.

\begin{table}[t]
	\centering
	\caption{Composite collision results with known transmitted bits}
	\label{tab:meas_composite_metrics}
	\footnotesize
	\setlength{\tabcolsep}{3pt}
	\renewcommand{\arraystretch}{1.25}
	\rowcolors{2}{white}{tblzebra}
	\begin{tabularx}{0.95\columnwidth}{@{}>{\centering\arraybackslash}p{0.35\columnwidth}>{\centering\arraybackslash}X>{\centering\arraybackslash}p{0.23\columnwidth}@{}}
		\rowcolor{tblhead}
		\hd{Case} & \hd{Tag count estimation accuracy} & \hd{Tag success rate} \\
		$R=2$ ($360$ composites) & $99.2\%$ & $100.0\%$ \\
		$R=3$ ($360$ composites) & $98.9\%$ & $99.4\%$ \\
		$R=4$ ($360$ composites) & $100.0\%$ & $95.6\%$ \\
	\end{tabularx}
	\vspace{-0.5cm}
\end{table}

The composite results show that SATR maintains high tag count estimation accuracy and tag success rate across all evaluated collision sizes. Since each recorded response contains receiver noise, these noise contributions are also combined, resulting in per tag SNRs of approximately $24$, $22$, and $21$ dB for $R=2$, $3$, and $4$. At the corresponding SNR values, Fig.~\ref{fig:satr_tag_success} gives tag success rates between $0.85$ and $0.98$, and the measured composite results broadly agree with this range. This evaluation also helps to interpret the lower RN16 agreement in the direct collision captures. In the composite case, the transmitted bits are known and the channel estimates are obtained from the same single tag recordings used to form each collision, allowing the BCJR reference to decode all composites correctly. In direct collisions, the transmitted bits are unavailable and the channels may differ from the single tag estimates because of mutual coupling between nearby tag antennas, so the agreement metric cannot distinguish SATR decoding errors from errors in the BCJR reference.

These measured decoding results evaluate signal recovery independently of the reader control loop. In a real-time reader, SATR inference and preprocessing must be completed within the $T_2$ interval, before the reader issues the next command, so latency remains a practical constraint. In addition, the full throughput gains of the multiple acknowledgment setting require protocol support for acknowledging more than one recovered RN16 response from a collided slot. Thus, the measurements validate the waveform-level recovery capability of SATR, while real-time integration with the reader protocol remains an important direction for practical deployment.

\section{Conclusion}
\label{sec:conclusion}

In this paper, we studied multi-tag collision recovery for passive UHF-RFID under the EPC Gen2 standard. We proposed Self-Attention Tag Recovery (SATR), a transformer-based decoding algorithm that recovers individual RN16 sequences directly from the sampled baseband I/Q waveform of a collided slot. SATR combines a waveform encoder that captures the I/Q structure and FM0 transitions with learned candidate tag representations to jointly estimate the collision size and recover the RN16 sequences without CSI or additional pilot symbols. We also analyzed the FSA throughput enabled by multi-tag recovery under single and multiple acknowledgment settings. Numerical results showed that SATR provides reliable collision recovery and the resulting FSA throughput approaches the theoretical limits of the corresponding recovery configurations. Measurements with commercial UHF-RFID tags further validated SATR for tag count estimation and RN16 recovery in collisions with up to four responding tags. These results indicate that the decoding performance of SATR approaches that of optimal detection without requiring the same side information, making learned physical layer decoding attractive when channel coefficients and signal parameters are difficult to estimate reliably.

Several directions remain for future work. Scaling the candidate output set to larger collisions could improve FSA throughput, while training SATR with channel and timing variations could strengthen robustness. Additional directions include evaluating SATR with moving tags and different tag types, integrating it with the reader control loop, and examining protocol support for multiple acknowledgments.

\bibliographystyle{IEEEtran}
\bibliography{ref}

\end{document}